\documentclass[longauth]{aa}

\usepackage{graphicx}
\usepackage{natbib}
\usepackage{scalerel}

\newcommand{\sersic}{S\'ersic\xspace}

\usepackage[table]{xcolor}

\usepackage{txfonts}
\usepackage[pdfencoding=auto,psdextra]{hyperref}
\hypersetup{
    colorlinks=true,
    linkcolor=blue,
    filecolor=magenta,      
    urlcolor=blue,
    citecolor=blue
}
\makeatletter
\renewcommand*\aa@pageof{, page \thepage{} of \pageref*{LastPage}}
\makeatother

\usepackage[utf8]{inputenc}

\usepackage[switch, modulo]{lineno}
\renewcommand{\linenumbers}[0]{}

\usepackage{euclid}

\usepackage{changes}

\begin{document}
%
%


   

%
\title{\Euclid preparation}
\subtitle{Refining input galaxy shape distributions for shear calibration simulations}      

\newcommand{\orcid}[1]{} 
\author{Euclid Collaboration: H.~Jansen\orcid{0009-0002-1332-7742}\thanks{\email{Henning.Jansen@uibk.ac.at}}\inst{\ref{aff1}}
\and N.~Martinet\orcid{0000-0003-2786-7790}\inst{\ref{aff2}}
\and S.~Grandis\orcid{0000-0002-4577-8217}\inst{\ref{aff1}}
\and H.~Hoekstra\orcid{0000-0002-0641-3231}\inst{\ref{aff3}}
\and S.-S.~Li\orcid{0000-0001-9952-7408}\inst{\ref{aff4},\ref{aff5}}
\and T.~Schrabback\orcid{0000-0002-6987-7834}\inst{\ref{aff1}}
\and G.~Congedo\orcid{0000-0003-2508-0046}\inst{\ref{aff6}}
\and B.~Csizi\orcid{0000-0003-3227-6581}\inst{\ref{aff1}}
\and F.~Kleinebreil\inst{\ref{aff1}}
\and G.~Mankar\orcid{0009-0001-3523-0813}\inst{\ref{aff1}}
\and N.~Zimmermann\inst{\ref{aff1}}
\and B.~Altieri\orcid{0000-0003-3936-0284}\inst{\ref{aff7}}
\and S.~Andreon\orcid{0000-0002-2041-8784}\inst{\ref{aff8}}
\and N.~Auricchio\orcid{0000-0003-4444-8651}\inst{\ref{aff9}}
\and C.~Baccigalupi\orcid{0000-0002-8211-1630}\inst{\ref{aff10},\ref{aff11},\ref{aff12},\ref{aff13}}
\and M.~Baldi\orcid{0000-0003-4145-1943}\inst{\ref{aff14},\ref{aff9},\ref{aff15}}
\and S.~Bardelli\orcid{0000-0002-8900-0298}\inst{\ref{aff9}}
\and P.~Battaglia\orcid{0000-0002-7337-5909}\inst{\ref{aff9}}
\and A.~Biviano\orcid{0000-0002-0857-0732}\inst{\ref{aff11},\ref{aff10}}
\and E.~Branchini\orcid{0000-0002-0808-6908}\inst{\ref{aff16},\ref{aff17},\ref{aff8}}
\and M.~Brescia\orcid{0000-0001-9506-5680}\inst{\ref{aff18},\ref{aff19}}
\and S.~Camera\orcid{0000-0003-3399-3574}\inst{\ref{aff20},\ref{aff21},\ref{aff22}}
\and V.~Capobianco\orcid{0000-0002-3309-7692}\inst{\ref{aff22}}
\and C.~Carbone\orcid{0000-0003-0125-3563}\inst{\ref{aff23}}
\and V.~F.~Cardone\inst{\ref{aff24},\ref{aff25}}
\and J.~Carretero\orcid{0000-0002-3130-0204}\inst{\ref{aff26},\ref{aff27}}
\and S.~Casas\orcid{0000-0002-4751-5138}\inst{\ref{aff28},\ref{aff29}}
\and F.~J.~Castander\orcid{0000-0001-7316-4573}\inst{\ref{aff30},\ref{aff31}}
\and M.~Castellano\orcid{0000-0001-9875-8263}\inst{\ref{aff24}}
\and G.~Castignani\orcid{0000-0001-6831-0687}\inst{\ref{aff9}}
\and S.~Cavuoti\orcid{0000-0002-3787-4196}\inst{\ref{aff19},\ref{aff32}}
\and A.~Cimatti\inst{\ref{aff33}}
\and C.~Colodro-Conde\inst{\ref{aff34}}
\and L.~Conversi\orcid{0000-0002-6710-8476}\inst{\ref{aff35},\ref{aff7}}
\and Y.~Copin\orcid{0000-0002-5317-7518}\inst{\ref{aff36}}
\and F.~Courbin\orcid{0000-0003-0758-6510}\inst{\ref{aff37},\ref{aff38},\ref{aff39}}
\and H.~M.~Courtois\orcid{0000-0003-0509-1776}\inst{\ref{aff40}}
\and M.~Cropper\orcid{0000-0003-4571-9468}\inst{\ref{aff41}}
\and H.~Degaudenzi\orcid{0000-0002-5887-6799}\inst{\ref{aff42}}
\and G.~De~Lucia\orcid{0000-0002-6220-9104}\inst{\ref{aff11}}
\and H.~Dole\orcid{0000-0002-9767-3839}\inst{\ref{aff43}}
\and F.~Dubath\orcid{0000-0002-6533-2810}\inst{\ref{aff42}}
\and X.~Dupac\inst{\ref{aff7}}
\and S.~Escoffier\orcid{0000-0002-2847-7498}\inst{\ref{aff44}}
\and M.~Farina\orcid{0000-0002-3089-7846}\inst{\ref{aff45}}
\and R.~Farinelli\inst{\ref{aff9}}
\and S.~Farrens\orcid{0000-0002-9594-9387}\inst{\ref{aff46}}
\and S.~Ferriol\inst{\ref{aff36}}
\and P.~Fosalba\orcid{0000-0002-1510-5214}\inst{\ref{aff31},\ref{aff30}}
\and S.~Fotopoulou\orcid{0000-0002-9686-254X}\inst{\ref{aff47}}
\and N.~Fourmanoit\orcid{0009-0005-6816-6925}\inst{\ref{aff44}}
\and M.~Frailis\orcid{0000-0002-7400-2135}\inst{\ref{aff11}}
\and E.~Franceschi\orcid{0000-0002-0585-6591}\inst{\ref{aff9}}
\and M.~Fumana\orcid{0000-0001-6787-5950}\inst{\ref{aff23}}
\and S.~Galeotta\orcid{0000-0002-3748-5115}\inst{\ref{aff11}}
\and K.~George\orcid{0000-0002-1734-8455}\inst{\ref{aff48}}
\and B.~Gillis\orcid{0000-0002-4478-1270}\inst{\ref{aff6}}
\and C.~Giocoli\orcid{0000-0002-9590-7961}\inst{\ref{aff9},\ref{aff15}}
\and J.~Gracia-Carpio\orcid{0000-0003-4689-3134}\inst{\ref{aff49}}
\and A.~Grazian\orcid{0000-0002-5688-0663}\inst{\ref{aff50}}
\and F.~Grupp\inst{\ref{aff49},\ref{aff51}}
\and S.~V.~H.~Haugan\orcid{0000-0001-9648-7260}\inst{\ref{aff52}}
\and W.~Holmes\orcid{0009-0007-8554-4646}\inst{\ref{aff53}}
\and I.~M.~Hook\orcid{0000-0002-2960-978X}\inst{\ref{aff54}}
\and F.~Hormuth\inst{\ref{aff55}}
\and A.~Hornstrup\orcid{0000-0002-3363-0936}\inst{\ref{aff56},\ref{aff57}}
\and K.~Jahnke\orcid{0000-0003-3804-2137}\inst{\ref{aff58}}
\and M.~Jhabvala\inst{\ref{aff59}}
\and B.~Joachimi\orcid{0000-0001-7494-1303}\inst{\ref{aff60}}
\and S.~Kermiche\orcid{0000-0002-0302-5735}\inst{\ref{aff44}}
\and A.~Kiessling\orcid{0000-0002-2590-1273}\inst{\ref{aff53}}
\and B.~Kubik\orcid{0009-0006-5823-4880}\inst{\ref{aff36}}
\and K.~Kuijken\orcid{0000-0002-3827-0175}\inst{\ref{aff3}}
\and M.~K\"ummel\orcid{0000-0003-2791-2117}\inst{\ref{aff51}}
\and M.~Kunz\orcid{0000-0002-3052-7394}\inst{\ref{aff61}}
\and H.~Kurki-Suonio\orcid{0000-0002-4618-3063}\inst{\ref{aff62},\ref{aff63}}
\and A.~M.~C.~Le~Brun\orcid{0000-0002-0936-4594}\inst{\ref{aff64}}
\and S.~Ligori\orcid{0000-0003-4172-4606}\inst{\ref{aff22}}
\and P.~B.~Lilje\orcid{0000-0003-4324-7794}\inst{\ref{aff52}}
\and V.~Lindholm\orcid{0000-0003-2317-5471}\inst{\ref{aff62},\ref{aff63}}
\and I.~Lloro\orcid{0000-0001-5966-1434}\inst{\ref{aff65}}
\and G.~Mainetti\orcid{0000-0003-2384-2377}\inst{\ref{aff66}}
\and O.~Mansutti\orcid{0000-0001-5758-4658}\inst{\ref{aff11}}
\and O.~Marggraf\orcid{0000-0001-7242-3852}\inst{\ref{aff67}}
\and M.~Martinelli\orcid{0000-0002-6943-7732}\inst{\ref{aff24},\ref{aff25}}
\and F.~Marulli\orcid{0000-0002-8850-0303}\inst{\ref{aff68},\ref{aff9},\ref{aff15}}
\and R.~J.~Massey\orcid{0000-0002-6085-3780}\inst{\ref{aff69}}
\and E.~Medinaceli\orcid{0000-0002-4040-7783}\inst{\ref{aff9}}
\and M.~Meneghetti\orcid{0000-0003-1225-7084}\inst{\ref{aff9},\ref{aff15}}
\and E.~Merlin\orcid{0000-0001-6870-8900}\inst{\ref{aff24}}
\and G.~Meylan\orcid{0000-0001-6503-0209}\inst{\ref{aff70}}
\and A.~Mora\orcid{0000-0002-1922-8529}\inst{\ref{aff71}}
\and M.~Moresco\orcid{0000-0002-7616-7136}\inst{\ref{aff68},\ref{aff9}}
\and B.~Morin\inst{\ref{aff46},\ref{aff72}}
\and L.~Moscardini\orcid{0000-0002-3473-6716}\inst{\ref{aff68},\ref{aff9},\ref{aff15}}
\and R.~Nakajima\orcid{0009-0009-1213-7040}\inst{\ref{aff67}}
\and C.~Neissner\orcid{0000-0001-8524-4968}\inst{\ref{aff73},\ref{aff27}}
\and S.-M.~Niemi\orcid{0009-0005-0247-0086}\inst{\ref{aff74}}
\and J.~W.~Nightingale\orcid{0000-0002-8987-7401}\inst{\ref{aff75}}
\and C.~Padilla\orcid{0000-0001-7951-0166}\inst{\ref{aff73}}
\and S.~Paltani\orcid{0000-0002-8108-9179}\inst{\ref{aff42}}
\and F.~Pasian\orcid{0000-0002-4869-3227}\inst{\ref{aff11}}
\and K.~Pedersen\inst{\ref{aff76}}
\and W.~J.~Percival\orcid{0000-0002-0644-5727}\inst{\ref{aff77},\ref{aff78},\ref{aff79}}
\and V.~Pettorino\orcid{0000-0002-4203-9320}\inst{\ref{aff74}}
\and S.~Pires\orcid{0000-0002-0249-2104}\inst{\ref{aff46}}
\and G.~Polenta\orcid{0000-0003-4067-9196}\inst{\ref{aff80}}
\and L.~A.~Popa\inst{\ref{aff81}}
\and F.~Raison\orcid{0000-0002-7819-6918}\inst{\ref{aff49}}
\and A.~Renzi\orcid{0000-0001-9856-1970}\inst{\ref{aff82},\ref{aff83},\ref{aff9}}
\and J.~Rhodes\orcid{0000-0002-4485-8549}\inst{\ref{aff53}}
\and G.~Riccio\inst{\ref{aff19}}
\and M.~Roncarelli\orcid{0000-0001-9587-7822}\inst{\ref{aff9}}
\and R.~Saglia\orcid{0000-0003-0378-7032}\inst{\ref{aff51},\ref{aff49}}
\and Z.~Sakr\orcid{0000-0002-4823-3757}\inst{\ref{aff84},\ref{aff85},\ref{aff86}}
\and D.~Sapone\orcid{0000-0001-7089-4503}\inst{\ref{aff87}}
\and P.~Schneider\orcid{0000-0001-8561-2679}\inst{\ref{aff67}}
\and A.~Secroun\orcid{0000-0003-0505-3710}\inst{\ref{aff44}}
\and E.~Sihvola\orcid{0000-0003-1804-7715}\inst{\ref{aff88}}
\and P.~Simon\inst{\ref{aff67}}
\and C.~Sirignano\orcid{0000-0002-0995-7146}\inst{\ref{aff82},\ref{aff83}}
\and G.~Sirri\orcid{0000-0003-2626-2853}\inst{\ref{aff15}}
\and L.~Stanco\orcid{0000-0002-9706-5104}\inst{\ref{aff83}}
\and P.~Tallada-Cresp\'{i}\orcid{0000-0002-1336-8328}\inst{\ref{aff26},\ref{aff27}}
\and A.~N.~Taylor\inst{\ref{aff6}}
\and I.~Tereno\orcid{0000-0002-4537-6218}\inst{\ref{aff89},\ref{aff90}}
\and N.~Tessore\orcid{0000-0002-9696-7931}\inst{\ref{aff41}}
\and S.~Toft\orcid{0000-0003-3631-7176}\inst{\ref{aff91},\ref{aff92}}
\and R.~Toledo-Moreo\orcid{0000-0002-2997-4859}\inst{\ref{aff93}}
\and F.~Torradeflot\orcid{0000-0003-1160-1517}\inst{\ref{aff27},\ref{aff26}}
\and I.~Tutusaus\orcid{0000-0002-3199-0399}\inst{\ref{aff30},\ref{aff31},\ref{aff85}}
\and J.~Valiviita\orcid{0000-0001-6225-3693}\inst{\ref{aff62},\ref{aff63}}
\and T.~Vassallo\orcid{0000-0001-6512-6358}\inst{\ref{aff11},\ref{aff48}}
\and G.~Verdoes~Kleijn\orcid{0000-0001-5803-2580}\inst{\ref{aff94}}
\and Y.~Wang\orcid{0000-0002-4749-2984}\inst{\ref{aff95}}
\and J.~Weller\orcid{0000-0002-8282-2010}\inst{\ref{aff51},\ref{aff49}}
\and G.~Zamorani\orcid{0000-0002-2318-301X}\inst{\ref{aff9}}
\and F.~M.~Zerbi\orcid{0000-0002-9996-973X}\inst{\ref{aff8}}
\and E.~Zucca\orcid{0000-0002-5845-8132}\inst{\ref{aff9}}
\and M.~Ballardini\orcid{0000-0003-4481-3559}\inst{\ref{aff96},\ref{aff97},\ref{aff9}}
\and A.~Boucaud\orcid{0000-0001-7387-2633}\inst{\ref{aff98}}
\and E.~Bozzo\orcid{0000-0002-8201-1525}\inst{\ref{aff42}}
\and C.~Burigana\orcid{0000-0002-3005-5796}\inst{\ref{aff99},\ref{aff100}}
\and R.~Cabanac\orcid{0000-0001-6679-2600}\inst{\ref{aff85}}
\and M.~Calabrese\orcid{0000-0002-2637-2422}\inst{\ref{aff101},\ref{aff23}}
\and A.~Cappi\inst{\ref{aff102},\ref{aff9}}
\and T.~Castro\orcid{0000-0002-6292-3228}\inst{\ref{aff11},\ref{aff12},\ref{aff10},\ref{aff103}}
\and J.~A.~Escartin~Vigo\inst{\ref{aff49}}
\and J.~Garc\'ia-Bellido\orcid{0000-0002-9370-8360}\inst{\ref{aff84}}
\and S.~Hemmati\orcid{0000-0003-2226-5395}\inst{\ref{aff95}}
\and E.~Jullo\orcid{0000-0002-9253-053X}\inst{\ref{aff2}}
\and J.~Macias-Perez\orcid{0000-0002-5385-2763}\inst{\ref{aff104}}
\and R.~Maoli\orcid{0000-0002-6065-3025}\inst{\ref{aff105},\ref{aff24}}
\and J.~Mart\'{i}n-Fleitas\orcid{0000-0002-8594-569X}\inst{\ref{aff106}}
\and N.~Mauri\orcid{0000-0001-8196-1548}\inst{\ref{aff33},\ref{aff15}}
\and R.~B.~Metcalf\orcid{0000-0003-3167-2574}\inst{\ref{aff68},\ref{aff9}}
\and P.~Monaco\orcid{0000-0003-2083-7564}\inst{\ref{aff107},\ref{aff11},\ref{aff12},\ref{aff10}}
\and A.~Pezzotta\orcid{0000-0003-0726-2268}\inst{\ref{aff8}}
\and M.~P\"ontinen\orcid{0000-0001-5442-2530}\inst{\ref{aff62}}
\and I.~Risso\orcid{0000-0003-2525-7761}\inst{\ref{aff8},\ref{aff17},\ref{aff16}}
\and V.~Scottez\orcid{0009-0008-3864-940X}\inst{\ref{aff108},\ref{aff109}}
\and M.~Sereno\orcid{0000-0003-0302-0325}\inst{\ref{aff9},\ref{aff15}}
\and M.~Tenti\orcid{0000-0002-4254-5901}\inst{\ref{aff15}}
\and M.~Tucci\inst{\ref{aff42}}
\and M.~Viel\orcid{0000-0002-2642-5707}\inst{\ref{aff10},\ref{aff11},\ref{aff13},\ref{aff12},\ref{aff103}}
\and M.~Wiesmann\orcid{0009-0000-8199-5860}\inst{\ref{aff52}}
\and Y.~Akrami\orcid{0000-0002-2407-7956}\inst{\ref{aff84},\ref{aff110}}
\and I.~T.~Andika\orcid{0000-0001-6102-9526}\inst{\ref{aff51}}
\and G.~Angora\orcid{0000-0002-0316-6562}\inst{\ref{aff19},\ref{aff96}}
\and S.~Anselmi\orcid{0000-0002-3579-9583}\inst{\ref{aff83},\ref{aff82},\ref{aff111}}
\and M.~Archidiacono\orcid{0000-0003-4952-9012}\inst{\ref{aff112},\ref{aff113}}
\and F.~Atrio-Barandela\orcid{0000-0002-2130-2513}\inst{\ref{aff114}}
\and L.~Baumont\orcid{0000-0002-1518-0150}\inst{\ref{aff107},\ref{aff11},\ref{aff10}}
\and L.~Bazzanini\orcid{0000-0003-0727-0137}\inst{\ref{aff96},\ref{aff9}}
\and D.~Bertacca\orcid{0000-0002-2490-7139}\inst{\ref{aff82},\ref{aff50},\ref{aff83}}
\and M.~Bethermin\orcid{0000-0002-3915-2015}\inst{\ref{aff115}}
\and F.~Beutler\orcid{0000-0003-0467-5438}\inst{\ref{aff6}}
\and A.~Blanchard\orcid{0000-0001-8555-9003}\inst{\ref{aff85}}
\and L.~Blot\orcid{0000-0002-9622-7167}\inst{\ref{aff116},\ref{aff64}}
\and M.~Bonici\orcid{0000-0002-8430-126X}\inst{\ref{aff77},\ref{aff23}}
\and S.~Borgani\orcid{0000-0001-6151-6439}\inst{\ref{aff107},\ref{aff10},\ref{aff11},\ref{aff12},\ref{aff103}}
\and M.~L.~Brown\orcid{0000-0002-0370-8077}\inst{\ref{aff117}}
\and S.~Bruton\orcid{0000-0002-6503-5218}\inst{\ref{aff118}}
\and A.~Calabro\orcid{0000-0003-2536-1614}\inst{\ref{aff24}}
\and B.~Camacho~Quevedo\orcid{0000-0002-8789-4232}\inst{\ref{aff10},\ref{aff13},\ref{aff11}}
\and F.~Caro\orcid{0009-0003-1053-0507}\inst{\ref{aff24}}
\and C.~S.~Carvalho\inst{\ref{aff90}}
\and F.~Cogato\orcid{0000-0003-4632-6113}\inst{\ref{aff68},\ref{aff9}}
\and A.~R.~Cooray\orcid{0000-0002-3892-0190}\inst{\ref{aff119}}
\and O.~Cucciati\orcid{0000-0002-9336-7551}\inst{\ref{aff9}}
\and J.~E.~Davies\orcid{0000-0002-5079-9098}\inst{\ref{aff58}}
\and T.~de~Boer\orcid{0000-0001-5486-2747}\inst{\ref{aff120}}
\and G.~Desprez\orcid{0000-0001-8325-1742}\inst{\ref{aff94}}
\and A.~D\'iaz-S\'anchez\orcid{0000-0003-0748-4768}\inst{\ref{aff121}}
\and S.~Di~Domizio\orcid{0000-0003-2863-5895}\inst{\ref{aff16},\ref{aff17}}
\and J.~M.~Diego\orcid{0000-0001-9065-3926}\inst{\ref{aff122}}
\and V.~Duret\orcid{0009-0009-0383-4960}\inst{\ref{aff44}}
\and M.~Y.~Elkhashab\orcid{0000-0001-9306-2603}\inst{\ref{aff11},\ref{aff12},\ref{aff107},\ref{aff10}}
\and Y.~Fang\orcid{0000-0002-0334-6950}\inst{\ref{aff51}}
\and A.~Finoguenov\orcid{0000-0002-4606-5403}\inst{\ref{aff62}}
\and A.~Franco\orcid{0000-0002-4761-366X}\inst{\ref{aff123},\ref{aff124},\ref{aff125}}
\and K.~Ganga\orcid{0000-0001-8159-8208}\inst{\ref{aff98}}
\and R.~Gavazzi\orcid{0000-0002-5540-6935}\inst{\ref{aff2},\ref{aff126}}
\and E.~Gaztanaga\orcid{0000-0001-9632-0815}\inst{\ref{aff30},\ref{aff31},\ref{aff127}}
\and F.~Giacomini\orcid{0000-0002-3129-2814}\inst{\ref{aff15}}
\and F.~Gianotti\orcid{0000-0003-4666-119X}\inst{\ref{aff9}}
\and G.~Gozaliasl\orcid{0000-0002-0236-919X}\inst{\ref{aff128},\ref{aff62}}
\and A.~Gruppuso\orcid{0000-0001-9272-5292}\inst{\ref{aff9},\ref{aff15}}
\and M.~Guidi\orcid{0000-0001-9408-1101}\inst{\ref{aff14},\ref{aff9}}
\and C.~M.~Gutierrez\orcid{0000-0001-7854-783X}\inst{\ref{aff34},\ref{aff129}}
\and A.~Hall\orcid{0000-0002-3139-8651}\inst{\ref{aff6}}
\and C.~Hern\'andez-Monteagudo\orcid{0000-0001-5471-9166}\inst{\ref{aff129},\ref{aff34}}
\and H.~Hildebrandt\orcid{0000-0002-9814-3338}\inst{\ref{aff130}}
\and J.~J.~E.~Kajava\orcid{0000-0002-3010-8333}\inst{\ref{aff131},\ref{aff132},\ref{aff133}}
\and Y.~Kang\orcid{0009-0000-8588-7250}\inst{\ref{aff42}}
\and V.~Kansal\orcid{0000-0002-4008-6078}\inst{\ref{aff134},\ref{aff135}}
\and D.~Karagiannis\orcid{0000-0002-4927-0816}\inst{\ref{aff96},\ref{aff136}}
\and K.~Kiiveri\orcid{0000-0002-3711-3346}\inst{\ref{aff88}}
\and J.~Kim\orcid{0000-0003-2776-2761}\inst{\ref{aff137}}
\and C.~C.~Kirkpatrick\inst{\ref{aff88}}
\and K.~Koyama\orcid{0000-0001-6727-6915}\inst{\ref{aff127}}
\and S.~Kruk\orcid{0000-0001-8010-8879}\inst{\ref{aff7}}
\and M.~C.~Lam\orcid{0000-0002-9347-2298}\inst{\ref{aff6}}
\and M.~Lattanzi\orcid{0000-0003-1059-2532}\inst{\ref{aff97}}
\and L.~Legrand\orcid{0000-0003-0610-5252}\inst{\ref{aff138},\ref{aff139}}
\and M.~Lembo\orcid{0000-0002-5271-5070}\inst{\ref{aff126}}
\and F.~Lepori\orcid{0009-0000-5061-7138}\inst{\ref{aff140}}
\and G.~Leroy\orcid{0009-0004-2523-4425}\inst{\ref{aff141},\ref{aff69}}
\and G.~F.~Lesci\orcid{0000-0002-4607-2830}\inst{\ref{aff68},\ref{aff9}}
\and J.~Lesgourgues\orcid{0000-0001-7627-353X}\inst{\ref{aff28}}
\and T.~I.~Liaudat\orcid{0000-0002-9104-314X}\inst{\ref{aff142}}
\and S.~J.~Liu\orcid{0000-0001-7680-2139}\inst{\ref{aff45}}
\and M.~Magliocchetti\orcid{0000-0001-9158-4838}\inst{\ref{aff45}}
\and F.~Mannucci\orcid{0000-0002-4803-2381}\inst{\ref{aff143}}
\and C.~J.~A.~P.~Martins\orcid{0000-0002-4886-9261}\inst{\ref{aff144},\ref{aff145}}
\and L.~Maurin\orcid{0000-0002-8406-0857}\inst{\ref{aff43}}
\and M.~Miluzio\inst{\ref{aff7},\ref{aff146}}
\and C.~Moretti\orcid{0000-0003-3314-8936}\inst{\ref{aff11},\ref{aff10},\ref{aff12}}
\and G.~Morgante\inst{\ref{aff9}}
\and S.~Nadathur\orcid{0000-0001-9070-3102}\inst{\ref{aff127}}
\and K.~Naidoo\orcid{0000-0002-9182-1802}\inst{\ref{aff127},\ref{aff58}}
\and A.~Navarro-Alsina\orcid{0000-0002-3173-2592}\inst{\ref{aff67}}
\and S.~Nesseris\orcid{0000-0002-0567-0324}\inst{\ref{aff84}}
\and F.~Pace\orcid{0000-0001-8039-0480}\inst{\ref{aff20},\ref{aff21},\ref{aff22}}
\and D.~Paoletti\orcid{0000-0003-4761-6147}\inst{\ref{aff9},\ref{aff100}}
\and F.~Passalacqua\orcid{0000-0002-8606-4093}\inst{\ref{aff82},\ref{aff83}}
\and K.~Paterson\orcid{0000-0001-8340-3486}\inst{\ref{aff58}}
\and L.~Patrizii\inst{\ref{aff15}}
\and C.~Pattison\orcid{0000-0003-3272-2617}\inst{\ref{aff127}}
\and A.~Pisani\orcid{0000-0002-6146-4437}\inst{\ref{aff44}}
\and D.~Potter\orcid{0000-0002-0757-5195}\inst{\ref{aff147}}
\and G.~W.~Pratt\inst{\ref{aff46}}
\and S.~Quai\orcid{0000-0002-0449-8163}\inst{\ref{aff68},\ref{aff9}}
\and M.~Radovich\orcid{0000-0002-3585-866X}\inst{\ref{aff50}}
\and K.~Rojas\orcid{0000-0003-1391-6854}\inst{\ref{aff148}}
\and W.~Roster\orcid{0000-0002-9149-6528}\inst{\ref{aff49}}
\and S.~Sacquegna\orcid{0000-0002-8433-6630}\inst{\ref{aff149}}
\and M.~Sahl\'en\orcid{0000-0003-0973-4804}\inst{\ref{aff150}}
\and D.~B.~Sanders\orcid{0000-0002-1233-9998}\inst{\ref{aff120}}
\and E.~Sarpa\orcid{0000-0002-1256-655X}\inst{\ref{aff11}}
\and A.~Schneider\orcid{0000-0001-7055-8104}\inst{\ref{aff147}}
\and D.~Sciotti\orcid{0009-0008-4519-2620}\inst{\ref{aff24},\ref{aff25}}
\and D.~Scognamiglio\orcid{0000-0001-8450-7885}\inst{\ref{aff151},\ref{aff53}}
\and E.~Sellentin\inst{\ref{aff152},\ref{aff3}}
\and L.~C.~Smith\orcid{0000-0002-3259-2771}\inst{\ref{aff153}}
\and E.~Soubrie\orcid{0000-0001-9295-1863}\inst{\ref{aff43}}
\and I.~Szapudi\orcid{0000-0003-2274-0301}\inst{\ref{aff120}}
\and K.~Tanidis\orcid{0000-0001-9843-5130}\inst{\ref{aff154}}
\and F.~Tarsitano\orcid{0000-0002-5919-0238}\inst{\ref{aff155},\ref{aff42}}
\and G.~Testera\orcid{0000-0003-2970-766X}\inst{\ref{aff17}}
\and M.~Tewes\orcid{0000-0002-1155-8689}\inst{\ref{aff67}}
\and R.~Teyssier\orcid{0000-0001-7689-0933}\inst{\ref{aff156}}
\and S.~Tosi\orcid{0000-0002-7275-9193}\inst{\ref{aff16},\ref{aff8},\ref{aff17}}
\and A.~Troja\orcid{0000-0003-0239-4595}\inst{\ref{aff11}}
\and C.~Uhlemann\orcid{0000-0001-7831-1579}\inst{\ref{aff157},\ref{aff75}}
\and C.~Valieri\inst{\ref{aff15}}
\and A.~Venhola\orcid{0000-0001-6071-4564}\inst{\ref{aff158}}
\and D.~Vergani\orcid{0000-0003-0898-2216}\inst{\ref{aff9}}
\and G.~Verza\orcid{0000-0002-1886-8348}\inst{\ref{aff159},\ref{aff160}}
\and E.~Vilenius\orcid{0000-0002-6184-7681}\inst{\ref{aff41}}
\and S.~Vinciguerra\orcid{0009-0005-4018-3184}\inst{\ref{aff2}}
\and M.~von~Wietersheim-Kramsta\orcid{0000-0003-4986-5091}\inst{\ref{aff69},\ref{aff141}}
\and N.~A.~Walton\orcid{0000-0003-3983-8778}\inst{\ref{aff153}}
\and A.~H.~Wright\orcid{0000-0001-7363-7932}\inst{\ref{aff130}}}
										   
\institute{Universit\"at Innsbruck, Institut f\"ur Astro- und Teilchenphysik, Technikerstr. 25/8, 6020 Innsbruck, Austria\label{aff1}
\and
Aix-Marseille Universit\'e, CNRS, CNES, LAM, Marseille, France\label{aff2}
\and
Leiden Observatory, Leiden University, Einsteinweg 55, 2333 CC Leiden, The Netherlands\label{aff3}
\and
Kavli Institute for Particle Astrophysics \& Cosmology (KIPAC), Stanford University, Stanford, CA 94305, USA\label{aff4}
\and
SLAC National Accelerator Laboratory, 2575 Sand Hill Road, Menlo Park, CA 94025, USA\label{aff5}
\and
Institute for Astronomy, University of Edinburgh, Royal Observatory, Blackford Hill, Edinburgh EH9 3HJ, UK\label{aff6}
\and
ESAC/ESA, Camino Bajo del Castillo, s/n., Urb. Villafranca del Castillo, 28692 Villanueva de la Ca\~nada, Madrid, Spain\label{aff7}
\and
INAF-Osservatorio Astronomico di Brera, Via Brera 28, 20122 Milano, Italy\label{aff8}
\and
INAF-Osservatorio di Astrofisica e Scienza dello Spazio di Bologna, Via Piero Gobetti 93/3, 40129 Bologna, Italy\label{aff9}
\and
IFPU, Institute for Fundamental Physics of the Universe, via Beirut 2, 34151 Trieste, Italy\label{aff10}
\and
INAF-Osservatorio Astronomico di Trieste, Via G. B. Tiepolo 11, 34143 Trieste, Italy\label{aff11}
\and
INFN, Sezione di Trieste, Via Valerio 2, 34127 Trieste TS, Italy\label{aff12}
\and
SISSA, International School for Advanced Studies, Via Bonomea 265, 34136 Trieste TS, Italy\label{aff13}
\and
Dipartimento di Fisica e Astronomia, Universit\`a di Bologna, Via Gobetti 93/2, 40129 Bologna, Italy\label{aff14}
\and
INFN-Sezione di Bologna, Viale Berti Pichat 6/2, 40127 Bologna, Italy\label{aff15}
\and
Dipartimento di Fisica, Universit\`a di Genova, Via Dodecaneso 33, 16146, Genova, Italy\label{aff16}
\and
INFN-Sezione di Genova, Via Dodecaneso 33, 16146, Genova, Italy\label{aff17}
\and
Department of Physics "E. Pancini", University Federico II, Via Cinthia 6, 80126, Napoli, Italy\label{aff18}
\and
INAF-Osservatorio Astronomico di Capodimonte, Via Moiariello 16, 80131 Napoli, Italy\label{aff19}
\and
Dipartimento di Fisica, Universit\`a degli Studi di Torino, Via P. Giuria 1, 10125 Torino, Italy\label{aff20}
\and
INFN-Sezione di Torino, Via P. Giuria 1, 10125 Torino, Italy\label{aff21}
\and
INAF-Osservatorio Astrofisico di Torino, Via Osservatorio 20, 10025 Pino Torinese (TO), Italy\label{aff22}
\and
INAF-IASF Milano, Via Alfonso Corti 12, 20133 Milano, Italy\label{aff23}
\and
INAF-Osservatorio Astronomico di Roma, Via Frascati 33, 00078 Monteporzio Catone, Italy\label{aff24}
\and
INFN-Sezione di Roma, Piazzale Aldo Moro, 2 - c/o Dipartimento di Fisica, Edificio G. Marconi, 00185 Roma, Italy\label{aff25}
\and
Centro de Investigaciones Energ\'eticas, Medioambientales y Tecnol\'ogicas (CIEMAT), Avenida Complutense 40, 28040 Madrid, Spain\label{aff26}
\and
Port d'Informaci\'{o} Cient\'{i}fica, Campus UAB, C. Albareda s/n, 08193 Bellaterra (Barcelona), Spain\label{aff27}
\and
Institute for Theoretical Particle Physics and Cosmology (TTK), RWTH Aachen University, 52056 Aachen, Germany\label{aff28}
\and
Deutsches Zentrum f\"ur Luft- und Raumfahrt e. V. (DLR), Linder H\"ohe, 51147 K\"oln, Germany\label{aff29}
\and
Institute of Space Sciences (ICE, CSIC), Campus UAB, Carrer de Can Magrans, s/n, 08193 Barcelona, Spain\label{aff30}
\and
Institut d'Estudis Espacials de Catalunya (IEEC),  Edifici RDIT, Campus UPC, 08860 Castelldefels, Barcelona, Spain\label{aff31}
\and
INFN section of Naples, Via Cinthia 6, 80126, Napoli, Italy\label{aff32}
\and
Dipartimento di Fisica e Astronomia "Augusto Righi" - Alma Mater Studiorum Universit\`a di Bologna, Viale Berti Pichat 6/2, 40127 Bologna, Italy\label{aff33}
\and
Instituto de Astrof\'{\i}sica de Canarias, E-38205 La Laguna, Tenerife, Spain\label{aff34}
\and
European Space Agency/ESRIN, Largo Galileo Galilei 1, 00044 Frascati, Roma, Italy\label{aff35}
\and
Universit\'e Claude Bernard Lyon 1, CNRS/IN2P3, IP2I Lyon, UMR 5822, Villeurbanne, F-69100, France\label{aff36}
\and
Institut de Ci\`{e}ncies del Cosmos (ICCUB), Universitat de Barcelona (IEEC-UB), Mart\'{i} i Franqu\`{e}s 1, 08028 Barcelona, Spain\label{aff37}
\and
Instituci\'o Catalana de Recerca i Estudis Avan\c{c}ats (ICREA), Passeig de Llu\'{\i}s Companys 23, 08010 Barcelona, Spain\label{aff38}
\and
Institut de Ciencies de l'Espai (IEEC-CSIC), Campus UAB, Carrer de Can Magrans, s/n Cerdanyola del Vall\'es, 08193 Barcelona, Spain\label{aff39}
\and
UCB Lyon 1, CNRS/IN2P3, IUF, IP2I Lyon, 4 rue Enrico Fermi, 69622 Villeurbanne, France\label{aff40}
\and
Mullard Space Science Laboratory, University College London, Holmbury St Mary, Dorking, Surrey RH5 6NT, UK\label{aff41}
\and
Department of Astronomy, University of Geneva, ch. d'Ecogia 16, 1290 Versoix, Switzerland\label{aff42}
\and
Universit\'e Paris-Saclay, CNRS, Institut d'astrophysique spatiale, 91405, Orsay, France\label{aff43}
\and
Aix-Marseille Universit\'e, CNRS/IN2P3, CPPM, Marseille, France\label{aff44}
\and
INAF-Istituto di Astrofisica e Planetologia Spaziali, via del Fosso del Cavaliere, 100, 00100 Roma, Italy\label{aff45}
\and
Universit\'e Paris-Saclay, Universit\'e Paris Cit\'e, CEA, CNRS, AIM, 91191, Gif-sur-Yvette, France\label{aff46}
\and
School of Physics, HH Wills Physics Laboratory, University of Bristol, Tyndall Avenue, Bristol, BS8 1TL, UK\label{aff47}
\and
University Observatory, LMU Faculty of Physics, Scheinerstr.~1, 81679 Munich, Germany\label{aff48}
\and
Max Planck Institute for Extraterrestrial Physics, Giessenbachstr. 1, 85748 Garching, Germany\label{aff49}
\and
INAF-Osservatorio Astronomico di Padova, Via dell'Osservatorio 5, 35122 Padova, Italy\label{aff50}
\and
Universit\"ats-Sternwarte M\"unchen, Fakult\"at f\"ur Physik, Ludwig-Maximilians-Universit\"at M\"unchen, Scheinerstr.~1, 81679 M\"unchen, Germany\label{aff51}
\and
Institute of Theoretical Astrophysics, University of Oslo, P.O. Box 1029 Blindern, 0315 Oslo, Norway\label{aff52}
\and
Jet Propulsion Laboratory, California Institute of Technology, 4800 Oak Grove Drive, Pasadena, CA, 91109, USA\label{aff53}
\and
Department of Physics, Lancaster University, Lancaster, LA1 4YB, UK\label{aff54}
\and
Felix Hormuth Engineering, Goethestr. 17, 69181 Leimen, Germany\label{aff55}
\and
Technical University of Denmark, Elektrovej 327, 2800 Kgs. Lyngby, Denmark\label{aff56}
\and
Cosmic Dawn Center (DAWN), Denmark\label{aff57}
\and
Max-Planck-Institut f\"ur Astronomie, K\"onigstuhl 17, 69117 Heidelberg, Germany\label{aff58}
\and
NASA Goddard Space Flight Center, Greenbelt, MD 20771, USA\label{aff59}
\and
Department of Physics and Astronomy, University College London, Gower Street, London WC1E 6BT, UK\label{aff60}
\and
Universit\'e de Gen\`eve, D\'epartement de Physique Th\'eorique and Centre for Astroparticle Physics, 24 quai Ernest-Ansermet, CH-1211 Gen\`eve 4, Switzerland\label{aff61}
\and
Department of Physics, P.O. Box 64, University of Helsinki, 00014 Helsinki, Finland\label{aff62}
\and
Helsinki Institute of Physics, Gustaf H{\"a}llstr{\"o}min katu 2, University of Helsinki, 00014 Helsinki, Finland\label{aff63}
\and
Laboratoire d'etude de l'Univers et des phenomenes eXtremes, Observatoire de Paris, Universit\'e PSL, Sorbonne Universit\'e, CNRS, 92190 Meudon, France\label{aff64}
\and
SKAO, Jodrell Bank, Lower Withington, Macclesfield SK11 9FT, UK\label{aff65}
\and
Centre de Calcul de l'IN2P3/CNRS, 21 avenue Pierre de Coubertin 69627 Villeurbanne Cedex, France\label{aff66}
\and
Universit\"at Bonn, Argelander-Institut f\"ur Astronomie, Auf dem H\"ugel 71, 53121 Bonn, Germany\label{aff67}
\and
Dipartimento di Fisica e Astronomia "Augusto Righi" - Alma Mater Studiorum Universit\`a di Bologna, via Piero Gobetti 93/2, 40129 Bologna, Italy\label{aff68}
\and
Department of Physics, Institute for Computational Cosmology, Durham University, South Road, Durham, DH1 3LE, UK\label{aff69}
\and
Institute of Physics, Laboratory of Astrophysics, Ecole Polytechnique F\'ed\'erale de Lausanne (EPFL), Observatoire de Sauverny, 1290 Versoix, Switzerland\label{aff70}
\and
Telespazio UK S.L. for European Space Agency (ESA), Camino bajo del Castillo, s/n, Urbanizacion Villafranca del Castillo, Villanueva de la Ca\~nada, 28692 Madrid, Spain\label{aff71}
\and
CEA Saclay, DFR/IRFU, Service d'Astrophysique, Bat. 709, 91191 Gif-sur-Yvette, France\label{aff72}
\and
Institut de F\'{i}sica d'Altes Energies (IFAE), The Barcelona Institute of Science and Technology, Campus UAB, 08193 Bellaterra (Barcelona), Spain\label{aff73}
\and
European Space Agency/ESTEC, Keplerlaan 1, 2201 AZ Noordwijk, The Netherlands\label{aff74}
\and
School of Mathematics, Statistics and Physics, Newcastle University, Herschel Building, Newcastle-upon-Tyne, NE1 7RU, UK\label{aff75}
\and
DARK, Niels Bohr Institute, University of Copenhagen, Jagtvej 155, 2200 Copenhagen, Denmark\label{aff76}
\and
Waterloo Centre for Astrophysics, University of Waterloo, Waterloo, Ontario N2L 3G1, Canada\label{aff77}
\and
Department of Physics and Astronomy, University of Waterloo, Waterloo, Ontario N2L 3G1, Canada\label{aff78}
\and
Perimeter Institute for Theoretical Physics, Waterloo, Ontario N2L 2Y5, Canada\label{aff79}
\and
Space Science Data Center, Italian Space Agency, via del Politecnico snc, 00133 Roma, Italy\label{aff80}
\and
Institute of Space Science, Str. Atomistilor, nr. 409 M\u{a}gurele, Ilfov, 077125, Romania\label{aff81}
\and
Dipartimento di Fisica e Astronomia "G. Galilei", Universit\`a di Padova, Via Marzolo 8, 35131 Padova, Italy\label{aff82}
\and
INFN-Padova, Via Marzolo 8, 35131 Padova, Italy\label{aff83}
\and
Instituto de F\'isica Te\'orica UAM-CSIC, Campus de Cantoblanco, 28049 Madrid, Spain\label{aff84}
\and
Institut de Recherche en Astrophysique et Plan\'etologie (IRAP), Universit\'e de Toulouse, CNRS, UPS, CNES, 14 Av. Edouard Belin, 31400 Toulouse, France\label{aff85}
\and
Universit\'e St Joseph; Faculty of Sciences, Beirut, Lebanon\label{aff86}
\and
Departamento de F\'isica, FCFM, Universidad de Chile, Blanco Encalada 2008, Santiago, Chile\label{aff87}
\and
Department of Physics and Helsinki Institute of Physics, Gustaf H\"allstr\"omin katu 2, University of Helsinki, 00014 Helsinki, Finland\label{aff88}
\and
Departamento de F\'isica, Faculdade de Ci\^encias, Universidade de Lisboa, Edif\'icio C8, Campo Grande, PT1749-016 Lisboa, Portugal\label{aff89}
\and
Instituto de Astrof\'isica e Ci\^encias do Espa\c{c}o, Faculdade de Ci\^encias, Universidade de Lisboa, Tapada da Ajuda, 1349-018 Lisboa, Portugal\label{aff90}
\and
Cosmic Dawn Center (DAWN)\label{aff91}
\and
Niels Bohr Institute, University of Copenhagen, Jagtvej 128, 2200 Copenhagen, Denmark\label{aff92}
\and
Universidad Polit\'ecnica de Cartagena, Departamento de Electr\'onica y Tecnolog\'ia de Computadoras,  Plaza del Hospital 1, 30202 Cartagena, Spain\label{aff93}
\and
Kapteyn Astronomical Institute, University of Groningen, PO Box 800, 9700 AV Groningen, The Netherlands\label{aff94}
\and
Caltech/IPAC, 1200 E. California Blvd., Pasadena, CA 91125, USA\label{aff95}
\and
Dipartimento di Fisica e Scienze della Terra, Universit\`a degli Studi di Ferrara, Via Giuseppe Saragat 1, 44122 Ferrara, Italy\label{aff96}
\and
Istituto Nazionale di Fisica Nucleare, Sezione di Ferrara, Via Giuseppe Saragat 1, 44122 Ferrara, Italy\label{aff97}
\and
Universit\'e Paris Cit\'e, CNRS, Astroparticule et Cosmologie, 75013 Paris, France\label{aff98}
\and
INAF, Istituto di Radioastronomia, Via Piero Gobetti 101, 40129 Bologna, Italy\label{aff99}
\and
INFN-Bologna, Via Irnerio 46, 40126 Bologna, Italy\label{aff100}
\and
Astronomical Observatory of the Autonomous Region of the Aosta Valley (OAVdA), Loc. Lignan 39, I-11020, Nus (Aosta Valley), Italy\label{aff101}
\and
Universit\'e C\^{o}te d'Azur, Observatoire de la C\^{o}te d'Azur, CNRS, Laboratoire Lagrange, Bd de l'Observatoire, CS 34229, 06304 Nice cedex 4, France\label{aff102}
\and
ICSC - Centro Nazionale di Ricerca in High Performance Computing, Big Data e Quantum Computing, Via Magnanelli 2, Bologna, Italy\label{aff103}
\and
Univ. Grenoble Alpes, CNRS, Grenoble INP, LPSC-IN2P3, 53, Avenue des Martyrs, 38000, Grenoble, France\label{aff104}
\and
Dipartimento di Fisica, Sapienza Universit\`a di Roma, Piazzale Aldo Moro 2, 00185 Roma, Italy\label{aff105}
\and
Aurora Technology for European Space Agency (ESA), Camino bajo del Castillo, s/n, Urbanizacion Villafranca del Castillo, Villanueva de la Ca\~nada, 28692 Madrid, Spain\label{aff106}
\and
Dipartimento di Fisica - Sezione di Astronomia, Universit\`a di Trieste, Via Tiepolo 11, 34131 Trieste, Italy\label{aff107}
\and
Institut d'Astrophysique de Paris, 98bis Boulevard Arago, 75014, Paris, France\label{aff108}
\and
ICL, Junia, Universit\'e Catholique de Lille, LITL, 59000 Lille, France\label{aff109}
\and
CERCA/ISO, Department of Physics, Case Western Reserve University, 10900 Euclid Avenue, Cleveland, OH 44106, USA\label{aff110}
\and
Laboratoire Univers et Th\'eorie, Observatoire de Paris, Universit\'e PSL, Universit\'e Paris Cit\'e, CNRS, 92190 Meudon, France\label{aff111}
\and
Dipartimento di Fisica "Aldo Pontremoli", Universit\`a degli Studi di Milano, Via Celoria 16, 20133 Milano, Italy\label{aff112}
\and
INFN-Sezione di Milano, Via Celoria 16, 20133 Milano, Italy\label{aff113}
\and
Departamento de F{\'\i}sica Fundamental. Universidad de Salamanca. Plaza de la Merced s/n. 37008 Salamanca, Spain\label{aff114}
\and
Universit\'e de Strasbourg, CNRS, Observatoire astronomique de Strasbourg, UMR 7550, 67000 Strasbourg, France\label{aff115}
\and
Center for Data-Driven Discovery, Kavli IPMU (WPI), UTIAS, The University of Tokyo, Kashiwa, Chiba 277-8583, Japan\label{aff116}
\and
Jodrell Bank Centre for Astrophysics, Department of Physics and Astronomy, University of Manchester, Oxford Road, Manchester M13 9PL, UK\label{aff117}
\and
California Institute of Technology, 1200 E California Blvd, Pasadena, CA 91125, USA\label{aff118}
\and
Department of Physics \& Astronomy, University of California Irvine, Irvine CA 92697, USA\label{aff119}
\and
Institute for Astronomy, University of Hawaii, 2680 Woodlawn Drive, Honolulu, HI 96822, USA\label{aff120}
\and
Departamento F\'isica Aplicada, Universidad Polit\'ecnica de Cartagena, Campus Muralla del Mar, 30202 Cartagena, Murcia, Spain\label{aff121}
\and
Instituto de F\'isica de Cantabria, Edificio Juan Jord\'a, Avenida de los Castros, 39005 Santander, Spain\label{aff122}
\and
INFN, Sezione di Lecce, Via per Arnesano, CP-193, 73100, Lecce, Italy\label{aff123}
\and
Department of Mathematics and Physics E. De Giorgi, University of Salento, Via per Arnesano, CP-I93, 73100, Lecce, Italy\label{aff124}
\and
INAF-Sezione di Lecce, c/o Dipartimento Matematica e Fisica, Via per Arnesano, 73100, Lecce, Italy\label{aff125}
\and
Institut d'Astrophysique de Paris, UMR 7095, CNRS, and Sorbonne Universit\'e, 98 bis boulevard Arago, 75014 Paris, France\label{aff126}
\and
Institute of Cosmology and Gravitation, University of Portsmouth, Portsmouth PO1 3FX, UK\label{aff127}
\and
Department of Computer Science, Aalto University, PO Box 15400, Espoo, FI-00 076, Finland\label{aff128}
\and
Universidad de La Laguna, Dpto. Astrof\'\i sica, E-38206 La Laguna, Tenerife, Spain\label{aff129}
\and
Ruhr University Bochum, Faculty of Physics and Astronomy, Astronomical Institute (AIRUB), German Centre for Cosmological Lensing (GCCL), 44780 Bochum, Germany\label{aff130}
\and
Department of Physics and Astronomy, Vesilinnantie 5, University of Turku, 20014 Turku, Finland\label{aff131}
\and
Finnish Centre for Astronomy with ESO (FINCA), Quantum, Vesilinnantie 5, University of Turku, 20014 Turku, Finland\label{aff132}
\and
Serco for European Space Agency (ESA), Camino bajo del Castillo, s/n, Urbanizacion Villafranca del Castillo, Villanueva de la Ca\~nada, 28692 Madrid, Spain\label{aff133}
\and
ARC Centre of Excellence for Dark Matter Particle Physics, Melbourne, Australia\label{aff134}
\and
Centre for Astrophysics \& Supercomputing, Swinburne University of Technology,  Hawthorn, Victoria 3122, Australia\label{aff135}
\and
Department of Physics and Astronomy, University of the Western Cape, Bellville, Cape Town, 7535, South Africa\label{aff136}
\and
Department of Physics, Oxford University, Keble Road, Oxford OX1 3RH, UK\label{aff137}
\and
DAMTP, Centre for Mathematical Sciences, Wilberforce Road, Cambridge CB3 0WA, UK\label{aff138}
\and
Kavli Institute for Cosmology Cambridge, Madingley Road, Cambridge, CB3 0HA, UK\label{aff139}
\and
Departement of Theoretical Physics, University of Geneva, Switzerland\label{aff140}
\and
Department of Physics, Centre for Extragalactic Astronomy, Durham University, South Road, Durham, DH1 3LE, UK\label{aff141}
\and
IRFU, CEA, Universit\'e Paris-Saclay 91191 Gif-sur-Yvette Cedex, France\label{aff142}
\and
INAF-Osservatorio Astrofisico di Arcetri, Largo E. Fermi 5, 50125, Firenze, Italy\label{aff143}
\and
Centro de Astrof\'{\i}sica da Universidade do Porto, Rua das Estrelas, 4150-762 Porto, Portugal\label{aff144}
\and
Instituto de Astrof\'isica e Ci\^encias do Espa\c{c}o, Universidade do Porto, CAUP, Rua das Estrelas, PT4150-762 Porto, Portugal\label{aff145}
\and
HE Space for European Space Agency (ESA), Camino bajo del Castillo, s/n, Urbanizacion Villafranca del Castillo, Villanueva de la Ca\~nada, 28692 Madrid, Spain\label{aff146}
\and
Department of Astrophysics, University of Zurich, Winterthurerstrasse 190, 8057 Zurich, Switzerland\label{aff147}
\and
University of Applied Sciences and Arts of Northwestern Switzerland, School of Computer Science, 5210 Windisch, Switzerland\label{aff148}
\and
INAF - Osservatorio Astronomico d'Abruzzo, Via Maggini, 64100, Teramo, Italy\label{aff149}
\and
Theoretical astrophysics, Department of Physics and Astronomy, Uppsala University, Box 516, 751 37 Uppsala, Sweden\label{aff150}
\and
Department of Physics, Duke University, Box 90305, Durham, NC 27708, USA\label{aff151}
\and
Mathematical Institute, University of Leiden, Einsteinweg 55, 2333 CA Leiden, The Netherlands\label{aff152}
\and
Institute of Astronomy, University of Cambridge, Madingley Road, Cambridge CB3 0HA, UK\label{aff153}
\and
Center for Astrophysics and Cosmology, University of Nova Gorica, Nova Gorica, Slovenia\label{aff154}
\and
Institute for Particle Physics and Astrophysics, Dept. of Physics, ETH Zurich, Wolfgang-Pauli-Strasse 27, 8093 Zurich, Switzerland\label{aff155}
\and
Department of Astrophysical Sciences, Peyton Hall, Princeton University, Princeton, NJ 08544, USA\label{aff156}
\and
Fakult\"at f\"ur Physik, Universit\"at Bielefeld, Postfach 100131, 33501 Bielefeld, Germany\label{aff157}
\and
Space physics and astronomy research unit, University of Oulu, Pentti Kaiteran katu 1, FI-90014 Oulu, Finland\label{aff158}
\and
International Centre for Theoretical Physics (ICTP), Strada Costiera 11, 34151 Trieste, Italy\label{aff159}
\and
Center for Computational Astrophysics, Flatiron Institute, 162 5th Avenue, 10010, New York, NY, USA\label{aff160}}      

%
%
\abstract{The Euclid Wide Survey (EWS) will cover the majority of the extragalactic sky with a resolution similar to the {\it Hubble} Space Telescope. This unprecedented data set will introduce a new era of precision cosmology. However, systematic effects need to be controlled better than ever. One of the sources of systematic uncertainties in weak gravitational lensing are biases introduced during the shear measurement. Determining these biases precisely allows the calibration of cosmological measurements to within \Euclid's required accuracy. The simulations that are used to determine such biases, need to resemble the real observations. In this work, we aim to learn distributions of galaxy shape parameters from real \Euclid data and use the new information to augment the morphological information in the Flagship galaxy mock catalogue. The morphology is extracted using single and double-\sersic model fits to the real data, for which we use \texttt{SourceXtractor++}. We train our pipeline on deep \Euclid observations of a field with rich auxiliary data and then use it to simulate EWS-like data. The input morphological distributions are tuned such that the measured distributions in the image simulations resemble the measured distributions in the real data. In these simulations we compare the multiplicative bias between the morphology from the Flagship catalogue, the trained single-\sersic morphology, and the trained double-\sersic morphology. We find that the image simulations with the updated morphology result in a percent-level change in the multiplicative shear bias compared to the original morphology from Flagship. This bias exceeds \Euclid's tight error budget by a factor of five and underlines the need for this work. Furthermore, we study the sensitivity of the multiplicative bias to key morphological parameters and show that our approach satisfies the requirements for the cosmology analysis with the first data release of \Euclid.}

%
%
%
%
   \titlerunning{\Euclid\/: Input morphology distributions for shear calibration simulations}
   \authorrunning{Euclid Collaboration: H.~Jansen et al.}
   
   \maketitle
%
%
%
%
   
\section{\label{sc:Intro}Introduction}
    The statistical power of the EWS \citep{Scaramella-EP1} conducted by ESA's \Euclid mission \citep{Laureijs11, EuclidSkyOverview} enables constraints of the cosmological model with unprecedented precision. To keep up with the statistical power, requirements on the knowledge of systematic effects are very tight \citep[see][]{Cropper_2013, Massey_2013}. One of the primary cosmology probes of \Euclid is weak gravitational lensing \citep[for a recent review see][]{prat_review}, which describes the distortion of the shapes of galaxies in the presence of foreground matter. This probe is systematically affected by the shear measurement bias, which is typically described as an additive bias and a multiplicative bias \citep[see][]{2005PhRvD..72d3503G, 2006MNRAS.366..101H, 2006MNRAS.368.1323H}. However, \cite{Hoekstra_21} shows that the additive component can be calibrated directly from the data itself. This is not the case for the multiplicative component. 
    The exact value of the shear measurement bias is highly sensitive to a plethora of different parameters in the simulations used to calibrate the bias \citep{Henk2017, Hernandez_2020, EP-Congedo, EP-Csizi}. Therefore, it is extremely important that the shear calibration of \Euclid takes into account all the features that are present in the real data. Recent weak lensing surveys like the Kilo-Degree Survey \citep[KiDS,][]{2013Msngr.154...44D}, the Dark Energy Survey \citep{DES}, and the Hyper Suprime-cam Subaru Strategic Program  \citep[HSC SSP,][]{HSC} make use of image simulations to calibrate the shear bias. Galaxies are simulated with a known true shear, and the bias can be determined by comparing the measured shear with the ground truth. The latest iterations of these image simulations \citep[see][]{2018MNRAS.481.3170M, 2019A&A...624A..92K, 2022MNRAS.509.3371M, SKiLLS} also ensure that the galaxy population in the image simulation looks statistically similar to that of the real sky. In particular \citet{SKiLLS} make use of the vine copula formalism \citep{VineCopulas} to capture the multidimensional morphology distribution from COSMOS observations and adopt it for the image simulations. 
    
    Here, we build on their work and adjust the input morphology of our simulations to match the observed \Euclid sky. We use the \Euclid Flagship simulation \citep{EuclidSkyFlagship} as a starting point. This $N$-body simulation with a large simulation box of $\SI{3600}{\hMpc}$ was designed specifically for \Euclid. The galaxy catalogue that comes with the simulation includes galaxy positions, redshifts, and also double-\sersic-profile \citep{sersic} morphologies, the latter of which we want to update based on observed \Euclid data. \citet{Henk2017} show that the multiplicative shear bias is particularly sensitive to the size and the ellipticity of galaxies. \citet{EP-Congedo} confirm that this bias is also sensitive to the bulge size when a two-component model is used. Therefore, we want to match sizes and ellipticities in particular very closely between image simulations and real data. Different studies \citep[see][]{Hoekstra_2015, Henk2017, Martinet-EP4} also highlight the importance of undetected faint galaxies for the accuracy of the shear bias estimation. The limiting magnitude for the nominal weak lensing sample of \Euclid is $\IE < 24.5$ \citep{Laureijs11}. Based on \cite{Martinet-EP4} we want to include galaxies at least two magnitudes fainter than that in the image simulations \citep[even fainter galaxies could be important as indicated by][]{EP-Congedo}. To get also their morphology right, we rely on deeper observations, which are obtained by \Euclid in the Euclid Auxiliary Fields and the Euclid Deep Survey \citep{EuclidSkyOverview}. For these fainter galaxies it is particularly important to get their sizes and clustering properties right, while the sensitivity to \sersic index and ellipticity is smaller. For this analysis we use the Cosmic Evolution Survey Deep Field \citep[COSMOS;][]{2007ApJS..172....1S} as observed by \Euclid, which has plenty of auxiliary data already taken. In this work, we use stacked images produced by the \Euclid MERging processing function \citep[see][]{Q1-TP004} with a median depth of $55\,000$ seconds. Hence, this data is exposed about 20 times longer than the nominal reference observation sequence (ROS) that is planned for the EWS \citep[see][]{Scaramella-EP1}. This corresponds to a factor 4.5 in the signal-to-noise ratio of sources. Ultimately we want to learn how galaxies look in these deeper data and then use this information to reproduce the observed morphology of galaxies in any observation of the EWS without further adjustments.  

    We introduce the weak lensing methods that we need for this work in Sect.\thinspace\ref{sc:Methods} and then proceed to summarizing the construction of the reference sample in Sect.\thinspace\ref{sc:Reference}. After that we describe the image simulations that we use to compare the reference sample to in Sect.\thinspace\ref{sc:Simulations} and then come to the details of the augmentation in Sect.\thinspace\ref{sc:Augmentation}. In Sect.\thinspace\ref{sc:sensitivity} we determine the sensitivity of the multiplicative bias to the input galaxy morphology in order to define our requirements. The outcome of the augmentation scheme is then shown in Sect.\thinspace\ref{sc:Results} together with the impact on the shear bias. Finally, we draw our conclusions in Sect.\thinspace\ref{sc:Conclusion}.

\section{\label{sc:Methods}Methods}
    \subsection{\label{sc:shear} Weak lensing shear}
        Gravitational lensing describes the deflection of light in the presence of a gravitational potential. For a general review about gravitational lensing see \citet{prat_review} and \citet{Schneider_2006}. Here we are only interested in the weak gravitational lensing, which occurs for larger line-of-sight separations between lens and source. In that case we can linearise the mapping between the lensed angular position and the angular position without lensing using the Jacobian matrix 
        \begin{equation}\label{eq:Jacobi}
            \mathit{A} = (1-\kappa)\begin{pmatrix} 1-g_1 & -g_2 \\ -g_2 & 1+g_1 \end{pmatrix}\,.
        \end{equation}
        This matrix is characterised by the two main quantities describing weak lensing: The convergence $\kappa$ and the (reduced) shear $g=g_1 + \mathrm{i}\,g_2$ (referred to only as `shear' hereafter). Hereby the convergence directly relates to the surface mass density along the line-of-sight and yields only an isotropic scaling of the apparent source size. 
        The shear is the anisotropic part of the Jacobian matrix and yields a distortion of the apparent shape of an object. For objects with confocal elliptical isophotes one can define ellipticity as
        \begin{equation}
            \epsilon = \frac{1-q}{1+q}\, \mathrm{e}^{2\mathrm{i}\varphi},
        \end{equation}
        with $q$ being the axis ratio, and $\varphi$ the orientation angle of the major axis. In this case one finds that 
        \begin{equation}
            g = \langle \epsilon_\mathrm{obs} \rangle\,,
        \end{equation}
        where $\epsilon_\mathrm{obs}$ denotes the observed ellipticity. This statement is based on the assumption that intrinsic galaxy orientations are random. In that case, averaging over a large sample of sources cancels out the intrinsic ellipticity of galaxies. In reality, the intrinsic alignment of galaxies due to the common formation history of spatially close galaxies, contaminates the signal and needs to be modeled. To determine the shear one needs a method to determine the ellipticity of an observed object. Over the past years different approaches have been developed, which we briefly highlight here. 

    \subsection{\label{sc:shear_measurement} Shear measurement}
        For practical reasons, the first attempts to determine the ellipticity of an object for a weak lensing analysis were moment based. Of the methods, the most widely used is \textsc{KSB+} \citep{KSB1, KSB2, KSB3}. It uses weighted second-order brightness moments 
        \begin{equation}
            Q_{ij} = \frac{\int I(\vec\theta)\,W(\vec\theta)\, \theta_i\,\theta_j\,\mathrm{d}^2\theta}{\int I(\vec\theta)\,W(\vec\theta)\,\mathrm{d}^2\theta}\,,
        \end{equation}
        where $\vec\theta$ denotes the image position relative to the object center (defined as where the weighted dipole vanishes), $I(\vec\theta)$ is the intensity at position $\vec\theta$, and $W(\vec\theta)$ is the weight function at position $\vec\theta$. Usually, one integrates over a square cutout around the source of interest. This estimate can also be corrected for the point-spread function (PSF) by measuring different moments of the PSF and galaxy image. It is a fast estimator, but can suffer from percent-level biases. A more recent approach is given by forward models, which fit the observed object by modelling a PSF-convolved light profile, and comparing it with the observation. The most recent iteration of this approach, which is used for the \Euclid analysis is called \textsc{LensMC} \citep{EP-Congedo}. It is a Bayesian forward model approach, where the ellipticity is estimated as the mean of the posterior distribution. In \textsc{LensMC} the posterior space is sampled with a Markov Chain Monte Carlo approach. Although highly runtime optimised already, this kind of approach is more computationally expensive than moment-based approaches. Still it has advantages like the possibility to fit close-by objects jointly. Another approach that is in development for \Euclid is \textsc{Metacalibration} \citep{Huff_2017, Sheldon_2017}. 
        \textsc{Metacalibration} is more a bias calibration scheme than a shear estimation algorithm. It can be used with either of the previously described approaches. \citet{yoon_25} demonstrate that \textsc{Metacalibration} reduces sensitivity to simulation specifics. It can be used directly on observed data and requires additional versions of an object to be built. With four additional versions of the object, where each is slightly sheared in one of two components, one can estimate the response matrix 
        \begin{equation}
            R_{ij} = \frac{\epsilon^+_i-\epsilon^-_j}{\Delta g_j}\,,
        \end{equation}
        where $\epsilon^+$ and $\epsilon^-$ denote the measured ellipticities of the slightly positively and negatively sheared versions of the same object. The shear difference between the two versions is denoted as $\Delta g_j$. 
        This response matrix describes how the shear measurement responds to a slight change of the shear. Usually the off-diagonal elements vanish on average and the diagonal elements tend to be equal \citep[see][]{yoon_25}. 
      
    \subsection{\label{sc:bias_meth}Shear bias}
        Estimating the shear of noisy galaxy images comes with many challenges. In general such an estimate is biased. For an accurate cosmological inference, it is therefore necessary to calibrate this bias and correct for it. The simplest and most popular bias model is a linear bias model as already used by early studies of weak gravitational lensing \citep[see][]{2005PhRvD..72d3503G, 2006MNRAS.366..101H, 2006MNRAS.368.1323H}. Recent works have shown that non-linearities can also become important for large shear values \citep[see][]{Jansen24, Shung_sheng_25}. The linear bias model, with the measured shear $\hat{g}$ and the true shear $g^\mathrm{true}$, follows
        \begin{equation}{\label{eq:linear_bias}}
            \hat{g}_i = (1+\mbias{i})\,g^\mathrm{true}_i + \cbias_i + \text{noise}\,,
        \end{equation}
        where $i=\{1,2\}$ denotes the two components of the complex shear. The multiplicative part $\mbias{i}$ yields an observed shear that is smaller or larger by a constant factor. We refer to this part of the bias as multiplicative shear bias. The additive part $\cbias_i$ adds an offset, which is independent of the true shear. This kind of bias is typically introduced by the PSF and is referred to as additive shear bias. Such biases can be introduced by a variety of processes. The dominant ones are the ellipticity measurement and the selection. Typically in the presence of noise the ellipticity measurement with the aforementioned methods is biased. Additionally, any selection of source galaxies, that has some implicit or explicit dependence on ellipticity, also introduces a bias. The extraction of sources from an image with \texttt{SExtractor} \citep[][]{SourceExtractor} tends to introduce a detection bias, since the detection probability is correlated with the ellipticity. Hence, even a perfect shape measurement method could still be biased by the selection \citep[e.g.][]{Hoekstra_et_al_2021}. 
        
        Each estimate of the shear is also intrinsically noisy, primarily due to the intrinsic galaxy ellipticities, and for noisy galaxy images additionally due to pixel noise. To mitigate the impact of this noise, the shear bias is typically estimated over a large sample of galaxies. With strategies like shape- and pixel noise cancellation it is possible to reduce the required simulation volume significantly as shown in \citet{Jansen24}. In this work we use \textsc{LensMC} with constant shears and a linear fit to determine the shear bias.

\section{\label{sc:Reference}Reference sample}
\begin{figure}
        \centering
        \resizebox{\hsize}{!}{\includegraphics{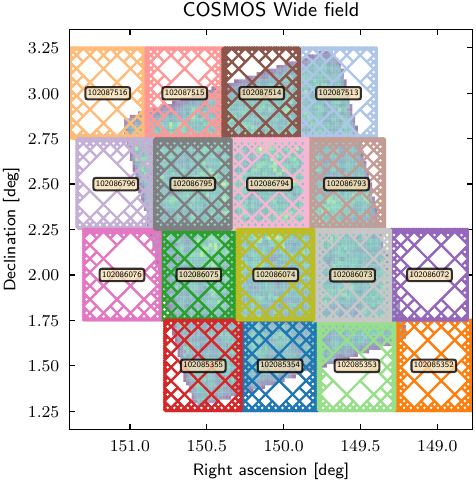}}
        \caption{The COSMOS field as observed by \Euclid with the corresponding MER tiles at EWS depth. The deep field covers about the same area with smaller tiles. We also show the \texttt{HEALpix} belonging to each tile and the underlying galaxy density in the background.}
        \label{fig:cosmos_coverage}
    \end{figure}

\begin{figure}
        \centering
        \resizebox{\hsize}{!}{\includegraphics{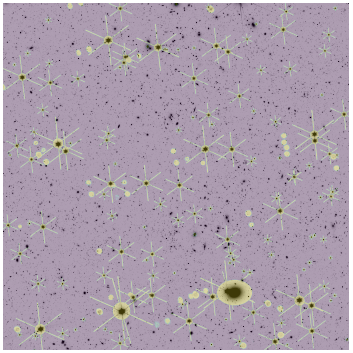}}
        \caption{Representative mask for one of the used tiles. We show the stellar masks, as well as the ghost masks, and custom masks for bright objects.}
        \label{fig:stellar_mask}
    \end{figure}

    We used real \Euclid observations as a reference sample to learn galaxy shape distributions. \citet{Henk2017} and \citet{Martinet-EP4} show that the calibration simulations need to include faint galaxies up to magnitude $28$ to get an accurate shear bias estimation. Therefore, we use deep observations of the COSMOS field \citep{2007ApJS..172....1S}, which was observed as one of \Euclid's calibration fields \citep[see][]{EuclidSkyOverview}, to learn galaxy morphology distributions. We combined 57 deep MER tiles \citep[see][]{Q1-TP004} to obtain a total of $1\,358\,726$ galaxies with morphological information. A MER tile is a stack of available observations in a pre-defined area of the sky. For the EWS these tiles cover $\SI{32}{\arcmin}\times\SI{32}{\arcmin}$, while the deep tiles are smaller measuring $\SI{17}{\arcmin}\times\SI{17}{\arcmin}$. Both of them have an overlap of $\SI{2}{\arcmin}$. Some of these tiles are not fully covered due to misalignment with the VIS focal plane. The effective area of our data covers $\SI{2.58}{\deg\squared}$. The coverage is visualized in Fig.\thinspace\ref{fig:cosmos_coverage}.
    For each tile, we also adopted a three-step masking process. In a first step, we applied the star masks that are provided by MER and doubled their size. That is because the automatic masking is designed for the EWS and does not yet account for the bigger apparent size of stars in the deeper stacks due to the extended faint light haloes of their images. The stars are masked by MER by scaling a template depending on the magnitude of the star and then rotating it according to the wcs information. Further we masked out the ghosts, that were flagged by the VIS processing function already. Lastly, we manually masked out ghosts that were not taken into account for by VIS. Also here the reason is that the masking accounts only for ghosts that would appear in the EWS, while in the deeper stacks also fainter ghosts become relevant. In Fig.\thinspace\ref{fig:stellar_mask} we show one representative tile and its masking. 
    
    We further applied an exposure time cut and only considered accumulated exposure times longer than $\SI{15}{\kilo\second}$ to be substantially deeper than the nominal wide survey. That cut and the applied masks further reduce the effective area to $\SI{2.17}{\deg\squared}$ and the number of sources to $1\,299\,721$. Additionally, we also measured a wide-like stack of the same field to validate that our learning method can be used to reproduce morphologies of galaxies in the wide survey. We generated a PSF model for each observation with \texttt{PSFex} \citep[][]{PSFex}, which models the PSF as a set of appropriate basis vectors. Hereby, we chose the sampling of the PSF model to be \ang{;;0.05} per pixel to deal better with the undersampled nature of the \Euclid PSF. To build the PSF model, we used stars from the {\it Gaia} star catalogue \citep[][]{GaiaDr3}, which overlapped with the field of view. The PSF model is an average over the whole tile and has no colour dependence.

    These PSF models were then used to generate model fits of galaxies in the observations. This process includes two steps. In a first step, we do the segmentation and detection of objects with \texttt{SExtractor} \citep[][]{SourceExtractor}. 
    \begin{table}[]
        \centering
        \caption{\texttt{SExtractor} parameters for the segmentation and detection.}
        \label{tab:sextractor_params}
        \begin{tabular}{l r}
        \hline\hline
          Parameter  &  Value\\ \hline 
           \texttt{DETECT\_MINAREA}  & 10 \\
           \texttt{DETECT\_THRESH} & 1.8 \\
           \texttt{FILTER\_NAME} & default.conv \\
           \texttt{DEBLEND\_NTHRESH} & 16 \\
           \texttt{DEBLEND\_MINCONT} & 0.1 \\
           \texttt{BACK\_SIZE} & 64 \\
           \texttt{BACK\_FILTERSIZE} & 5 \\
           \hline
        \end{tabular}
    \end{table}
    We list the key configuration parameters for this initial step in Table \ref{tab:sextractor_params}. The convolution filter is the standard one of \texttt{SExtractor}. It is a $3\times3$ filter with a FWHM of two pixels.  
    Then this pre-detected catalogue is passed to \texttt{SourceXtractor++} \citep[][]{Sepp2, Sepp1} to do the model fitting, since this software allows for more control. This software was already successfully applied to real data sets \citep[see][]{Bretonniere-EP13, Shuntov_25, EROLensData, Schrabback25}. 
    \begin{table}
        \centering
        \caption{Fitted parameters of the single-\sersic model.}
        \label{tab:sepp_modelfit}
        \begin{tabular}{l l c}
            \hline\hline
            Parameter & Initial value & Allowed range \\ \hline
            $x$ & \texttt{X\_IMAGE} & ($x_\mathrm{i} - r_\mathrm{i}$, $x_\mathrm{i} + r_\mathrm{i}$) \\
            $y$ & \texttt{Y\_IMAGE} & ($y_\mathrm{i} - r_\mathrm{i}$, $y_\mathrm{i} + r_\mathrm{i}$) \\
            flux & \texttt{FLUX\_AUTO} & free \\
            $r$ & \texttt{A\_IMAGE} & ($0.01\,r_\mathrm{i}$, $2\,r_\mathrm{i}$) \\
            $q$ & $\texttt{ELONGATION}^{-1}$ & (0.1, 1.0) \\
            angle & \texttt{THETA\_IMAGE} & free \\
            $\nSersic$ & 2.0 & (0.3, 6.2)  \\ \hline
        \end{tabular}
        \tablefoot{We use the subscript i to denote initial values, which can be read off in the initial column. The initial values themselves refer to the output parameters of the detection and segmentation run with \texttt{SExtractor}.}
    \end{table}
    The model we fit with all the parameters used can be seen in Table \ref{tab:sepp_modelfit}. 
    To constrain the ellipticity of the fitted object, we introduce parameters that are derived from the fitted parameters. Those are the absolute value of the ellipticity $|\epsilon|$, and the two ellipticity components
    \begin{align}
        \epsilon_1 &= |\epsilon|\,\cos(2\,\alpha) \,, \\
        \epsilon_2 &= |\epsilon|\,\sin(2\,\alpha)\,,
    \end{align}
    where $\alpha$ denotes the orientation angle of the major axis.
    
    We also control the box size in which the fit is done by passing the semi-major axis of \texttt{SExtractor} as a source size option to \texttt{SourceXtractor++}. To properly take care of close and blended objects, we group sources with a friends-of-friends algorithm using \texttt{pyfof}.\footnote{\url{https://github.com/simongibbons/pyfof}} We used a linking length of \ang{;;2}. Then the group indices are passed to \texttt{SourceXtractor++} for a joint fitting of associated sources. 
    \begin{figure}
        \centering
        \resizebox{\hsize}{!}{\includegraphics{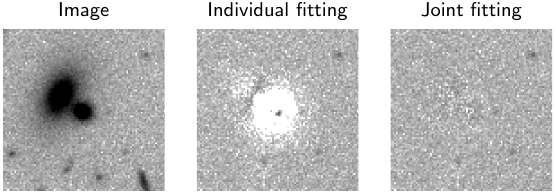}}
        \caption{Difference between joint and individual fitting. We notice that the residuals for a joint fit look like noise, while the individual fitting has clear structure. The fainter galaxy is over-corrected, because flux of the brighter one is being associated with it.}
        \label{fig:grouped_fits}
    \end{figure}
    In Fig.\thinspace\ref{fig:grouped_fits} we show an example of the advantage that this joint fitting has. The residuals and therefore the quality of the fit can be improved a lot.
    \begin{figure*}
        \centering
        \resizebox{\hsize}{!}{\includegraphics{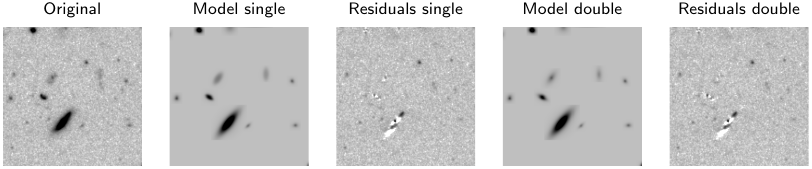}}
        \caption{Generic example of model fitting on a $150\times150$ pixel cutout. Stars are not fitted for. Both residuals look consistent with noise at the modeled positions indicating that the fitting works well.}
        \label{fig:model_fit_example}
    \end{figure*}
    An example of the complete fitting procedure can be seen in Fig.\thinspace\ref{fig:model_fit_example}. We can see the original image together with the models and the residuals. A few observations can be made here. First of all, not all sources that one can identify by eye are being modelled. This is mainly due to the detection settings in Table \ref{tab:sextractor_params} being rather conservative, where only sources with 10 connected pixels exceeding the background by $1.8\,\sigma$ are detected and propagated to the model fitting. 
    Additionally we see remaining substructure in the residuals especially for the brightest objects. That is to be expected since close-by galaxies often have resolved substructure like spiral arms, which are not captured by the single-\sersic model fit. 
    Overall, the fitted light profiles in the model look reasonable and therefore we proceeded with this scheme.  

    Since the masks cover only the brightest stars in the field, our sample is still contaminated by stars of fainter magnitudes. We performed a star-galaxy separation based on the measured half-light radius, which we found to separate the populations best. We describe the procedure in Appendix \ref{app:s_g}. 

    \subsection{Double-\sersic fit}
        To test the sensitivity of our workflow to the choice of the fitted model, we also fit a double-\sersic model to the observed data, which is a more complex parametric model. The model parameters are described in Table \ref{tab:sepp_modelfit_double}. 
        \begin{table}
            \centering
            \caption{Fitted parameters of the double-\sersic model.}
            \label{tab:sepp_modelfit_double}
            \begin{tabular}{l l c}
                \hline\hline
                Parameter & Initial value & Allowed range \\ \hline
                $x$ & \texttt{X\_IMAGE} & ($x_\mathrm{i} - r_\mathrm{i}$, $x_\mathrm{i} + r_\mathrm{i}$) \\
                $y$ & \texttt{Y\_IMAGE} & ($y_\mathrm{i} - r_\mathrm{i}$, $y_\mathrm{i} + r_\mathrm{i}$) \\
                flux & \texttt{FLUX\_AUTO} & free \\
                bulge fraction & 0.5 & (0, 1) \\
                $r_\mathrm{disc}$ & \texttt{A\_IMAGE} & ($0.01\,r_\mathrm{i}$, $2\,r_\mathrm{i}$)  \\
                $r_\mathrm{bulge}$ & 0.15\,\texttt{A\_IMAGE} & ($0.01\,r_\mathrm{i}$, $2\,r_\mathrm{i}$) \\
                $q_\mathrm{disc}$ & $\texttt{ELONGATION}^{-1}$ & (0.1, 1.0) \\
                $q_\mathrm{bulge}$ & $\texttt{ELONGATION}^{-1}$ & (0.1, 1.0) \\
                angle & \texttt{THETA\_IMAGE} & free \\
                $\nSersic^\mathrm{bulge}$ & 4.0 & (0.3, 6.2)  \\
                $\nSersic^\mathrm{disc}$ & 1.0 & fixed \\ \hline
            \end{tabular}
            \tablefoot{We use the subscript $\mathrm{i}$ to denote initial values, which can be read off in the initial column. The initial values themselves refer to the output parameters of the detection and segmentation run with \texttt{SExtractor}.}
        \end{table}
        To constrain the degrees of freedom we fix the light profile of the disc to be exponential. The reduced-$\chi^2$ distributions of the two models look very similar. Hence, the majority of the sources is fitted well with either model. With this fitting we have a single- and a double-\sersic fit for each detected object and can apply our augmentation scheme to both versions.

    \subsection{Redshift information}
        We complement our morphological fits with high-quality photometric redshifts from COSMOS2020 \citep{Weaver_2022} to capture the morphology--redshift dependence. Each of our sources gets assigned a photometric redshift based on a nearest neighbour matching. We consider as successful matches only those sources that have a neighbour within $\ang{;;0.5}$ in COSMOS2020. As a point estimate for the redshift we used the column \texttt{lp\_zPDF} generated with the \texttt{LePhare} code \citep{lephare1, lephare2}. 
        
        Due to masks in the COSMOS2020 catalogue, we do not find a match for every source. We expect no correlation between masked areas and morphology such that this is not problematic. Still, the fraction of sources with photometric redshifts starts decreasing further from magnitude 25 on. Therefore, we used the redshift information in the copulae only until this magnitude and generate the fainter magnitudes without redshift information. When we drew from copulae trained with redshift information, we conditioned the draw also on the redshift given by the Flagship catalogue. 
        
\section{\label{sc:Simulations}Image simulations}
    In order to verify the learned morphology distribution from the reference sample, we need to generate image simulations on which we can execute the same measurement procedure that we used to generate the reference sample. Those simulations built on the ones outlined in \cite{Jansen24} and therefore we only provide a short description of those here. In general the image simulations are based on \texttt{Galsim} \citep[][]{galsim}, which is a python library designed to generate image simulations of astronomical instruments. 
    \subsection{Galaxies}
        We generated galaxies with two simple parametric models. As our fiducial simulation, we rendered images of the Flagship galaxy catalogue, which was obtained from CosmoHub \citep{CosmoHub2, CosmoHub1}. The light distribution in this catalogue follows double-\sersic profiles, where bulge and disc have independent \sersic distributions. 

        One of our trained models, also uses double-\sersic profiles to model the light distribution. Therefore, each of its parameters exists also in the Flagship catalogue. For a second model, we simplified the light distribution for the morphology-learning from real data to a single-\sersic profile, which makes the model fit more stable. In all cases, we took the positions, magnitudes, and redshifts of the galaxies directly from the Flagship catalogue, since we expect those properties to be correctly modelled. 
        
    \subsection{PSF}
        The \sersic model is then convolved with the PSF and drawn on the pixel grid. We use directly the PSF that was modelled on the real data as described in Sect.\thinspace\ref{sc:Reference}. Hence the PSF model is not chromatic and is representative of the PSF for an average stellar spectral energy distribution (SED). \citet{cypriano_10}, \citet{eriksen_18}, and lately \citet{Schrabback25} discuss that the average SED of a galaxy is different from the one of stars. For this work, we expect the PSF model inaccuracy to be a second-order effect and therefore do not adopt a more complex treatment of the PSF. We use a constant PSF for each simulated tile, which is selected uniformly among the average PSF models of the reference data (Sect.~\ref{sc:Reference}).
        
    \subsection{Stars}
        We also used the aforementioned PSF model to simulate stars. We used the publicly available tool \texttt{TRILEGAL} \citep[see][]{Trilegal} via their online interface to generate VIS magnitudes of a stellar population at the sky position of our reference data.\footnote{\url{http://stev.oapd.inaf.it/cgi-bin/trilegal}} We generated $\SI{2.5}{\deg\squared}$ around the central position of the COSMOS field up to a limiting VIS magnitude of 28 using the default settings of \texttt{TRILEGAL}. Then we randomly drew a fraction of stars from that, which corresponds to the fraction of the tile area compared to the full $\SI{2.5}{\deg\squared}$. 
        \begin{figure}
            \centering
            \resizebox{\hsize}{!}{\includegraphics{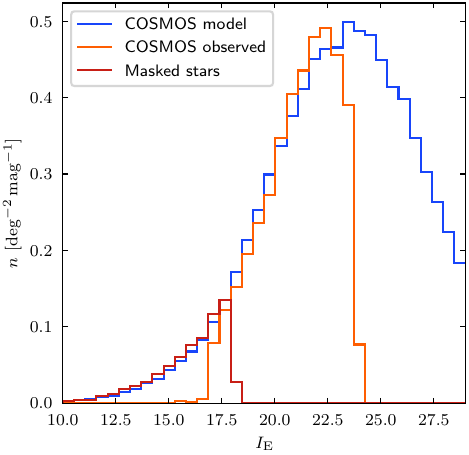}}
            \caption{Simulated stellar densities in the COSMOS field compared to the observed ones and the masked stars. The observed distribution was determined using $\texttt{POINT\_LIKE\_PROB}>0.8$ as a cut for the MER catalogue.}
            \label{fig:stellar_densities}
        \end{figure}
        The stellar densities in the COSMOS field can be seen in Fig.\thinspace\ref{fig:stellar_densities}. For reference we also show the observed stellar distribution given by a $\texttt{POINT\_LIKE\_PROB}>0.8$ cut on the MER catalogue. It is evident that both distributions agree very well. The observed distribution does not cover the full magnitude range, because very bright stars are saturated and masked. To illustrate the masking, we also plot the distribution of the masked stellar sources. The $\texttt{POINT\_LIKE\_PROB}$ classifier of MER is not designed to work at the faint end of the magnitude distribution, where observed sizes of galaxies tend to overlap with the size of the PSF. We kept track of the stellar positions that we simulated and used them later for the PSF modelling with \texttt{PSFex} as we did on the real data. 

    \subsection{Matching depth}
\begin{figure}
            \centering
            \resizebox{\hsize}{!}{\includegraphics{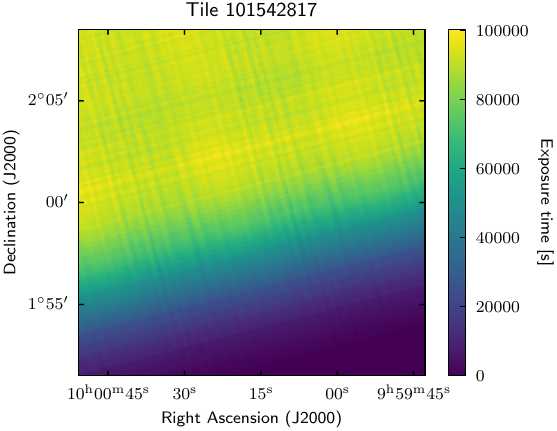}}
            \caption{Exposure time variation of one of the used deep MER tiles. One can see a strong gradient in the exposure time towards the upper part of the image, where more VIS observations overlap.}
            \label{fig:exp_time_selfcal}
        \end{figure}

        The images that we generated have a size of $4128\times4128$ pixels, where the outer 64 pixels on each side are left empty to minimise border effects. Therefore one simulated image is representative for one of the 36 CCDs of the VIS instrument \citep[see][]{EuclidSkyVIS}. 
        \begin{figure}
            \centering
            \resizebox{\hsize}{!}{\includegraphics{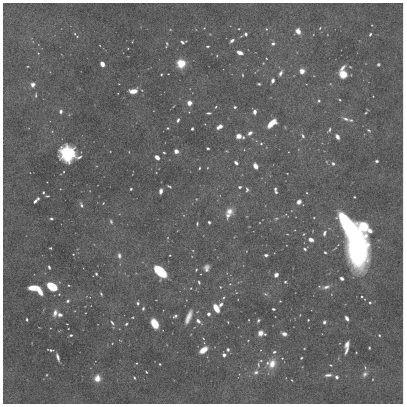}}
            \caption{Representative simulated $1128\times1128$ pixel tile of the COSMOS field in a logarithmic gray scaling. The bright almost perfectly round sources are stars, since the \texttt{PSFex} model only captures the inner part and no additional features like diffraction spikes. The outer 64 pixels in each direction do not contain any simulated sources to avoid border effects.}
            \label{fig:example_image_sim}
        \end{figure}
        In Fig.\thinspace\ref{fig:example_image_sim} we show an example of such a simulated image, which is only $1128\times1128$ pixel wide in this case for better visibility. To match the depth of the real observations that we use as a reference sample, we take the zero-point information from the header of the mosaics. The exposure time is not constant over the field of view due to the stacked tiles not being perfectly aligned with the original VIS exposures. As an example, we show this for one of the used tiles of the COSMOS field in Fig.\thinspace\ref{fig:exp_time_selfcal}. 
        Here we see that the upper left corner is exposed longer than the lower right corner. To account for that, we simulated our tiles with an exposure time distribution that matches the one of the reference field. Each individual tile has a constant exposure time, but the exposure time distribution over all tiles matches the one of the reference field. 
        \begin{figure}
            \centering
            \resizebox{\hsize}{!}{\includegraphics{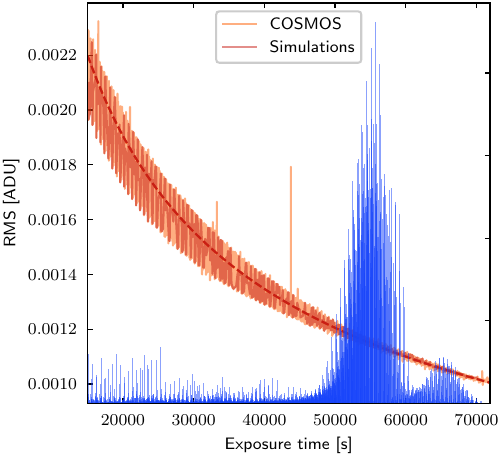}}
            \caption{Distribution of exposure times and background values over the field of view of the mosaics. The blue histogram shows the distribution of exposure times, while the lines show the behaviour of the background RMS given the exposure time. Dashed lines highlight the fitted behaviour for the RMS. }
            \label{fig:exposure_time_hist}
        \end{figure}
        For the noise we built a model given the actual MER stacked images and their RMS maps. From the RMS maps we can read off what the variation in the background should be at a certain exposure time. During the stacking process of the individual exposures correlated noise was introduced due to a bilinear resampling. Additionally, imperfection in the spatially dependent background subtraction can introduce noise correlations on large scales. The MER RMS map accounts for this additional correlated noise. The single-pixel RMS estimated by \texttt{SExtractor}, on the other hand, underestimates the true noise level in the image. To account for this correlated noise in the image simulations, we used \texttt{NaMaster} \citep{namaster} to fit the angular power spectrum of the noise in the MER stacks using the segmentation map as a mask. Then we generated random noise fields given the power spectrum and scaled them to the variance needed at a certain exposure time. We found that the RMS, which accounts for correlated noise (given by the MER RMS map), is typically about $1.5$ times higher than the single pixel RMS determined by \texttt{SExtractor}. Therefore we calculated the true RMS of each simulated tile and increased the detection threshold by the factor that it was higher than the \texttt{SExtractor} estimate to simulate the impact that the RMS map had as a weight map for the detection in the real data. This increased detection threshold also leads to an underestimation of the fitting uncertainties of \texttt{SourceXtractor++}. We do not use the estimated uncertainties in our analysis and therefore do not need to take this into account. In Fig.\thinspace\ref{fig:exposure_time_hist} we show the distribution of exposure times and their corresponding background values determined by reading of the RMS map in the real data. We fit a functional behaviour of 
        \begin{equation}
            \sigma_\mathrm{bg}(t) = \frac{a}{\sqrt{t}}\,,    
        \end{equation}
        to the data, where $t$ denotes the exposure time. We recover $a_\mathrm{\texttt{COSMOS}} = \SI{0.2695}{ADU\;\second^{1/2}}$ and $a_\mathrm{simulations}=\SI{0.2694}{ADU\;\second^{1/2}}$, which shows that we can reproduce the depth of the observed data well. In addition to the background noise, we also added the shot noise coming from the sources themselves. 
        On these image simulations we then measure morphological parameters in the same way as we do for real data. We describe this procedure in the next section.

\section{\label{sc:Augmentation}Augmentation}
\begin{figure}
        \centering
        \resizebox{\hsize}{!}{\includegraphics{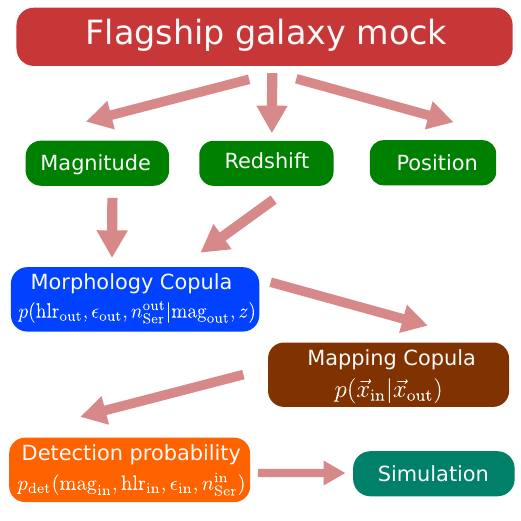}}
        \caption{Flagship augmentation scheme. We start from the Flagship galaxy mock and augment the morphology using two copulae and one dense neural network. The output of this scheme is the refined morphology for a galaxy at the requested magnitude and redshift, and a probability with which it should be injected in the image simulation.}
        \label{fig:augmentation_scheme}
    \end{figure}

    In this work we use the Flagship galaxy mock catalogue \citep{EuclidSkyFlagship} as a starting point to augment the morphology. The augmentation scheme is depicted in Fig.\thinspace\ref{fig:augmentation_scheme}.
    From the Flagship catalogue we employ the magnitude (calculated from the \texttt{euclid\_vis} flux), redshift, and position information for each object. The position and the redshift of the object is left unmodified since we assume the clustering in the simulation to be accurate. Starting from the magnitude and the redshift, we then augment the morphology. The three steps, which are taken from a flagship object to our simulations are described in the following. Since we saw in Sect.\thinspace\ref{sc:Simulations} that the exposure time can vary within a reference tile, all of these three steps are additionally conditioned on the exposure time. 
    \subsection{Morphology copula}
        Following \citet{SKiLLS} we learn the multi-dimensional distribution of morphology parameters from real data that is described in Sect.\thinspace\ref{sc:Reference} using vine copulae \citep{VineCopulas}. Copulae can be used to encode the interdependence of two random variables. In general a joint density of two variables $f(x_1, x_2)$ can be written as 
        \begin{equation}
            f(x_1, x_2) = f(x_1)\, f(x_2|x_1)\,.    
        \end{equation}
        Using copulae we can write 
        \begin{equation}
            f(x_1 \mid x_2) = c_{1,2}\!\left[F(x_1),F(x_2)\right]\, f(x_1)\,,
        \end{equation}
        where $F(x_1)$ and $F(x_2)$ are the one dimensional cumulative distribution functions and $c_{1,2}[a, b]$ denotes the copula. 
        In two dimensions many analytical copulae exist, such as the Gaussian copula. This is not the case any more in higher dimensions. Therefore one can build vines of bivariate copulae to still be able to model high-dimensional interdependence. 
        
        For the copula learning we use the publicly available package \texttt{pyvinecopulib}.\footnote{\url{https://github.com/vinecopulib/pyvinecopulib}} We build the vine in such a way that we can draw conditionally based on magnitude and redshift, which are dictated by the Flagship mock. The individual bivariate copulae are automatically chosen from all implemented possibilities to maximise the likelihood. To achieve higher stability of the copula learning, we train the copulae in magnitude bins. Otherwise the learned distribution would be completely dominated by the properties of the faintest galaxies, since they are the dominant population. We chose a binning with a spacing of $0.25$ magnitudes with an overlap of $0.025$ magnitudes to avoid border effects in the copulae. Additionally, we required a minimum number of a thousand galaxies to fall within the bin. Otherwise the bin was merged with the next one. Therefore, the binning is sparser for the brightest galaxies.  
        
        We start the augmentation in every case with the copula that was fitted on the observed single-\sersic-model parameters half-light radius, axis ratio, and \sersic index. Then we use the mapping copula to map the output single-\sersic parameters either to input single-\sersic or input double-\sersic parameters. This implementation is chosen because we only fit a single-\sersic model to the simulated data, which we compare with the fitted single-\sersic model of the observed data for the further analysis.

        Without any priors on the ellipticity, the fitting produces ellipticity spikes at the extreme values for faint galaxies due to the very uncertain fit. In practice, these galaxies are also too faint for a shape measurement and will not enter the cosmological analysis. Nevertheless, this behaviour is challenging for the copula to capture. Therefore, we pre-processed the ellipticity distributions as described in Appendix \ref{app:ellipticity} for a more stable training. 
        
    \subsection{Mapping copula}
\begin{figure*}
            \centering
            \resizebox{\hsize}{!}{\includegraphics{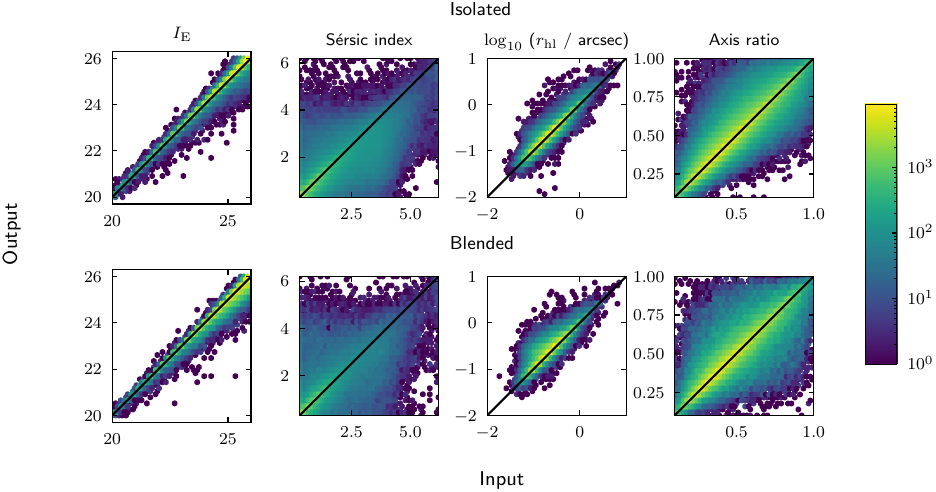}}
            \caption{Comparison between input galaxy parameters and recovered galaxy parameters from the image simulations. The half-light radius $r_\mathrm{hl}$ in arcsecond is displayed logarithmically here. The upper row shows isolated sources defined as having no neighbour within \ang{;;1}. The lower row indicates blended sources as those that have a neighbour within that radius. While most objects follow the identity line, there is also asymmetric scattering especially for magnitude and half-light radius.}
            \label{fig:mapping}
        \end{figure*}

        So far we only have learned the morphological parameters that we would like our measured galaxies to have. Still, we do not know if the image simulation maps its input to the measured output quantities. Many intermediate steps like the binning of the profile in pixels and the convolution with the PSF can potentially alter the distributions. We compare the morphological parameters in the input- and output catalogues by matching both catalogues. We match sources within a circle of $\ang{;;0.5}$ and take the closest one in magnitude space if there are multiple possible matches. In Fig.\thinspace\ref{fig:mapping} one can see the impact of the image simulation. 
        Because of the blending, we find strong trends in magnitude and the size of the objects. We did a rough separation between isolated and blended sources in Fig.\thinspace\ref{fig:mapping}, defining blended galaxies as those having a neighbour within \ang{;;1}. The faintest galaxies tend to scatter up in magnitude due to blending with brighter sources. At the same time small sources appear larger, because the outskirts of their profiles get additional flux from nearby neighbours. To invert this impact of blending, we train 8-dimensional vine copulae on the comparison between our input- and output catalogues. We fit again a vine of bivariate copulae, such that a conditional draw on the output properties is possible. For each object we want to know the morphological parameters that we need to inject into the simulation to acquire the desired measured morphological distribution of the population. This training is done with the same magnitude binning as the initial morphology copula. 
        
        For the double-\sersic fits, we did not compare directly input and output values of the same parameter. Instead, we compared input double-\sersic parameters with output single-\sersic parameters. We could then conditionally draw based on the output parameters to obtain the double-\sersic model that needs to be ingested in the simulation. In this way this copula both encompasses the mapping of the simulation and the mapping between the two parametric models. 
    \subsection{Incompleteness treatment}
        We face the problem of learning an unknown underlying ground truth distribution of morphological parameters from what we can detect and measure in the data. Especially at the faint end of the magnitude distribution we have a bias towards round and more peaked objects, since they distribute their limited flux over fewer pixels. Therefore, if we would just simulate all galaxies in the Flagship catalogue with the properties we learned from the observed sample, we will re-detect too many galaxies. To tackle this problem, we develop an incompleteness treatment which is as follows. 
        
        \subsubsection{Theory}\label{subsec:Theory}
            The morphology distribution that we can measure on the real data is incomplete. Thus, when we want to reproduce this distribution we need a treatment of this incompleteness. We formulate the problem as follows 
            \begin{equation}
                \underbrace{N_\mathrm{target}(m, \epsilon, r_\mathrm{h}, n)}_{\substack{\mathrm{Measured}}} = \underbrace{p(\mathrm{det}|m, \epsilon, r_\mathrm{h}, n)}_{\substack{\mathrm{Network}}} \, \underbrace{N_\mathrm{true}(m, \epsilon, r_\mathrm{h}, n)}_{\substack{\mathrm{Unknown}}}\,,
            \end{equation}
            where $m$ is the magnitude, $\epsilon$ the ellipticity, $r_\mathrm{h}$ the half-light radius, and $\nSersic$ the \sersic index. We do not know the true underlying distribution $N_\mathrm{true}$, but we can try to find a distribution $N_\mathrm{input}$ such that 
            \begin{equation}
                N_\mathrm{sim} = p(\mathrm{det}|m, \epsilon, r_\mathrm{h}, n)\, N_\mathrm{input}(m, \epsilon, r_\mathrm{h},n) = N_\mathrm{target}
            \end{equation}
            holds for the measured number counts $N_\mathrm{sim}$ from the simulation.
            The target number counts are measured from the real data. The detection probability $p(\mathrm{det}|m, \epsilon, r_\mathrm{h}, n)$, is determined by training a classifier on the outcome of simulations, where we can check if we detect a galaxy that we inserted in the first place. We describe this classifier in more detail in Sect.\thinspace\ref{subsec:det_prob}. The unknowns here are the input number counts that we need to put into the simulation to reproduce the target distribution as an outcome. 
            
            We want to build the $N_\mathrm{input}$ based on the Flagship catalogue to start from accurate clustering. Therefore, we can decompose $N_\mathrm{input}$ into 
                \begin{equation}
                    N_\mathrm{input}(m, \epsilon, r_\mathrm{h}, n) = N_\mathrm{Flagship}(m, \epsilon, r_\mathrm{h}, n)\, F(m, \epsilon, r_\mathrm{h}, n)\,,
                \end{equation}
            where $F(m, \epsilon, r_\mathrm{h},n )$ denotes a weight factor for the Flagship distribution. This weight factor is given by 
                \begin{equation}
                    F(m, \epsilon, r_\mathrm{h}, n) = \underbrace{\frac{N_\mathrm{target}(m, \epsilon, r_\mathrm{h}, n)}{N_\mathrm{Flagship}(m, \epsilon, r_\mathrm{h}, n)}}_{\substack{\mathrm{Downsampling}}}\, \underbrace{\frac{1}{p(\mathrm{det}|m, \epsilon, r_\mathrm{h},n)}}_{\substack{\mathrm{Boosting}}}\,.
                \end{equation}
            The first factor essentially equalises the magnitude distributions of Flagship and the measured distribution. Since we are incomplete, we will not detect every faint galaxy that we put into the simulations though. Thus we need to additionally boost the distribution depending on morphological parameters and the magnitude. So far the downsampling also depends on the morphology. Utilizing the copulae, we can remove the morphology dependence.
            
            We decompose the Flagship distribution as
                \begin{equation}
                    N_\mathrm{Flagship}(m, \epsilon, r_\mathrm{h}, n) = N_\mathrm{Flagship}(m)\, p_\mathrm{cop}(\epsilon, r_\mathrm{h}, n | m)\,. 
                \end{equation}
            Since the copula was trained on the target sample, we can say
                \begin{align}
                    N_\mathrm{target}(m, \epsilon, r_\mathrm{h}, n) &= p_\mathrm{target}(\epsilon, r_\mathrm{h}, n | m) \, N_\mathrm{target}(m) \nonumber \\  &\approx p_\mathrm{cop}(\epsilon, r_\mathrm{h}, n | m)\, N_\mathrm{target}(m) \,,
                \end{align}
            where $p_\mathrm{target}$ denotes the conditional probability of morphological parameters given the magnitude in the target sample, while $p_\mathrm{cop}$ denotes the same conditional probability of the trained copula.
            Hence, the expression for the weight factor simplifies to 
                \begin{equation}\label{eq:weight_factor}
                    F(m, \epsilon, r_\mathrm{h}, n) \approx \frac{N_\mathrm{target}(m)}{N_\mathrm{Flagship}(m)}\, \frac{1}{p(\mathrm{det}|m, \epsilon, r_\mathrm{h}, n)}\,.
                \end{equation}
            Given this weight factor, we can determine for every galaxy an injection probability with which it should be included in the simulation. With our setup we find this injection probability to be between $80\%$ and $100\%$ with some variation due to cosmic variance at the bright end and detection bias on the faint end. If we then re-detect on the simulation, the detection probability cancels out and we find 
                \begin{align}
                    N_\mathrm{sim}(m, \epsilon, r_\mathrm{h}, n) &= p(\mathrm{det}|m, \epsilon, r_\mathrm{h}, n)\, N_\mathrm{input}(m, \epsilon, r_\mathrm{h}, n) \nonumber \\ 
                     &= N_\mathrm{Flagship}(m, \epsilon, r_\mathrm{h},n)\, \frac{N_\mathrm{target}(m)}{N_\mathrm{Flagship}(m)} \nonumber \\
                     &= N_\mathrm{target}(m)\, p_\mathrm{cop}(\epsilon, r_\mathrm{h}, n | m) \nonumber \\
                     &\approx N_\mathrm{target}(m, \epsilon, r_\mathrm{h},n)\,,
                \end{align}
            which reproduces the target distribution. 

        \subsubsection{Magnitude weighting}
            The histogram equalization described by $N_\mathrm{target} / N_\mathrm{Flagship}$ also depends on the exposure time. Only in the parts of the image with the longest exposure time we can recover galaxies as faint as magnitude 28. To account for this in the simulations, we split the magnitude distribution into exposure time bins. 
            \begin{figure}
                \centering
                \resizebox{\hsize}{!}{\includegraphics{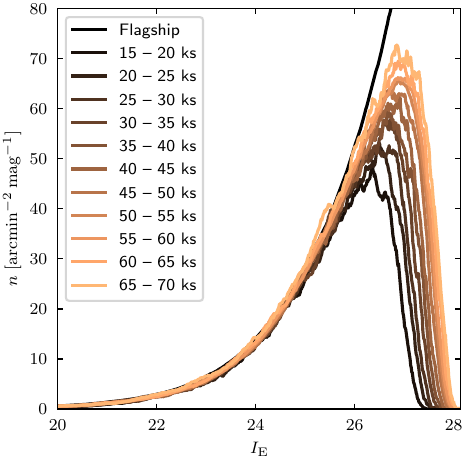}}
                \caption{Magnitude distribution of the real data for different exposure time ranges. The solid lines are kernel density estimates of the magnitude distribution in a given exposure time bin. As expected the longer we expose the more faint galaxies can be recovered.}
                \label{fig:magnitude_weighting}
            \end{figure}
            In Fig.\thinspace\ref{fig:magnitude_weighting} we show the observed magnitude distributions in the exposure time bins. As expected the depth of the image varies by about one magnitude depending on the exposure time. If we simulate a shallow part of the real data, we therefore do not need to include the faintest galaxy population. The solid lines indicate a kernel density estimation (KDE) of the magnitude distribution in each exposure time bin. Using the KDE allows us to have a continuous estimate of the density over the whole magnitude range. To calculate the weight factor described by Eq.\thinspace\eqref{eq:weight_factor}, we interpolate between the KDEs to get the number density of galaxies  $n_\mathrm{target}$ at a given exposure time. The weight is then determined as $n_\mathrm{target} / n_\mathrm{Flagship}$, where $n_\mathrm{Flagship}$ is the KDE of the Flagship magnitude distribution evaluated at a given magnitude.  
            
        \subsubsection{Detection probability}\label{subsec:det_prob}
\begin{figure}
                \centering
                \resizebox{\hsize}{!}{\includegraphics{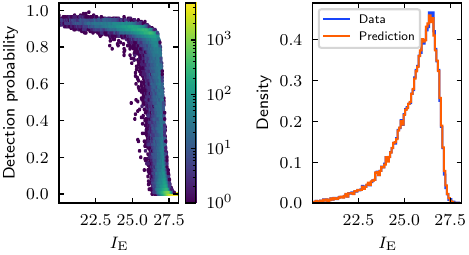}}
                \caption{Detection probability validation. The left panel shows the behavior of the detection probability with magnitude. In the right panel we show the magnitude histogram with the detection probability as weights.}
                \label{fig:network_validation}
            \end{figure}

            We want to model the detection probability of a galaxy given its magnitude and morphological parameters using a simple neural network. To build the training set, we do a simulation including all galaxies in the Flagship galaxy mock catalogue with morphological parameters assigned by the copulae. To define a re-detection we utilize the same matching scheme outlined before. The network is composed of an input layer with six inputs (magnitude, ellipticity, half-light radius, \sersic index, exposure time, and background RMS) for the single-\sersic model and nine inputs (magnitude, bulge axis ratio, disc axis ratio, half-light radius disc, half-light radius bulge, bulge \sersic index, bulge fraction, exposure time, and background RMS) for the double-\sersic model. This feeds into a dense layer with 32 neurons, an additional hidden layer with 16 neurons, and a sigmoid activated output layer. We utilize a binary cross-entropy loss function to train the model. The output of the model is then a continuous detection probability between $0$ and $1$.  
            In the right panel of Fig.\thinspace\ref{fig:network_validation} we show that the classifier is indeed well trainable and reproduces the training magnitude histogram. We also find that all other weighted histograms are well reproduced.  
            In the left panel of Fig.\thinspace\ref{fig:network_validation} we show the learned detection probability as a function of magnitude on the validation sample. We only trained for galaxies fainter than 20th magnitude, since we expect every brighter galaxy to be detected if it is not blended. The profile is shaped as expected for a detection curve. Therefore, we conclude that our detection probability network is suited to be used in the formalism described in the previous subsection. 

\section{\label{sc:sensitivity}{Sensitivity requirements}}
    To judge the capabilities of our augmentation scheme, we need to translate the requirements on the multiplicative bias uncertainty into requirements on key properties of the input parameter distributions. Therefore, we utilize the image simulations described in Sect.\thinspace\ref{sc:Simulations} to determine the multiplicative bias for a reference simulation and for slightly perturbed simulations. For the sensitivity analysis we simulated the images with the depth of the EWS. We used large shear values of up to $0.2$ together with shape- and pixel noise cancellation to maximize the efficiency. This allowed us to determine the sensitivity with an area of only $\SI{1.5}{\deg\squared}$ ($\times 4$ with both noise cancellations) per simulation. Furthermore, we kept the same noise realization for the reference simulation and the perturbed simulations to isolate the impact of the morphology.
    For the shear measurement, we utilize \textsc{LensMC} \citep{EP-Congedo}, which is one of the main shear measurements methods for \Euclid. To speed up the measurement, we used the PSF model that was used for the simulations without any oversampling. All galaxies with an observed magnitude between $20.5$ and $24.5$ are included in the analysis. The constraint on the bright end is chosen arbitrarily and excludes only a few very bright and local galaxies that would carry no shear in reality.
    
    The two most important morphological parameters are the ellipticity and the size of an object. Furthermore, we studied also the \sersic index. We perturb the mean (via an absolute shift) and the standard deviation (via a scaling of the distance to the mean) of the morphological parameters. For the ellipticity distribution, we additionally studied the impact of a kurtosis change. For the kurtosis, we followed an iterative approach. We transformed the original random variable $x$ via
    \begin{equation}
        y = x\,|x|^a\,,
    \end{equation}
    where $y$ is the transformed variable and $a$ is a free parameter to be determined. After each transformation, we scale $y$ back to the original mean and standard deviation via 
    \begin{align}
        y_\mathrm{norm} &= (y - \bar{y}) / \sigma_y\,, \\
        \hat{y} &= y_\mathrm{norm}\, \sigma_x + \bar{x}\,,
    \end{align}
    where $y_\mathrm{norm}$ is the deviation from the mean normalized to the standard deviation, and $\hat{y}$ describes the transformed distribution with the original mean and standard deviation. Then we measured the kurtosis of $\hat{y}$ and iterated until the difference between the measured and the target kurtosis was smaller than $0.01$. 
    This guarantees that only one of these properties is modified at a time. For each property we generated two perturbed simulations to have three points available for a linear fit (together with the reference simulation). In total we end up with the reference simulation and 14 perturbed simulations. Following \citet{Henk2017}, we assume $\sigma_m < 5\times 10^{-4}$ as our error budget for the multiplicative bias $\mbias{}$. This budget refers to the final mission requirements at \Euclid's third data release. Accounting for the expected increase in the number of galaxies from the first to the third data release, the requirement for the first data release is $\sigma_m < 1.4\times 10^{-3}$. In the following we work with the final requirements, but the DR1 requirements can be directly inferred by multiplying final requirements with $2.8$ (assuming a DR3 with $\SI{15000}{deg\squared}$ and a DR1 with $\SI{1900}{deg\squared}$). Hence, if the final requirements would allow for a $1\%$ shift in a certain parameter, the first data release still allows for $2.8\%$. With this error margin and the reference simulation, we can determine how far we can deviate from the reference simulation while recovering the correct multiplicative bias. 
    \begin{figure}
        \centering
        \resizebox{\hsize}{!}{\includegraphics{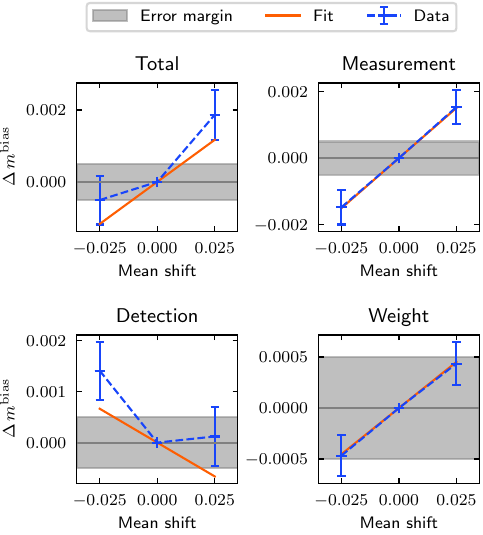}}
        \caption{Sensitivity determination for a mean shift in the absolute value of the ellipticity $\epsilon$. We show here the residual bias with respect to a reference simulation. The measurement bias arises solely from the shear measurement code (\textsc{LensMC} in this case), while the detection bias originates from a detection preference that correlates with the shear signal. The weight bias accounts for the difference between a weighted and an unweighted estimate of the bias.}
        \label{fig:ellipticity_shift}
    \end{figure}

    In Fig.\thinspace\ref{fig:ellipticity_shift} we illustrate the procedure for a mean shift of the ellipticity distribution. We force a linear fit to go through the reference point and then use the fitted slope to determine the required accuracy of the perturbed quantity. We split up the multiplicative bias in different contributions to see which one is most sensitive to the perturbation. The total bias was determined by using all galaxies and their shear weights. For the measurement bias we used exclusively galaxies that could be measured in all four realizations of shape- and pixel noise cancellation without their shear weights. This ensures that the intrinsic ellipticity vanishes on average and only the measurement itself leads to a bias. The detection bias is estimated from the difference between the unweighted total bias and the measurement bias. Finally for the weight bias we estimated the difference between the weighted- and the unweighted measurement bias. In the case shown here, the mean shift in the ellipticity causes a major change in the detection bias, while the measurement bias is largely unaffected. Following this procedure, we determined the sensitivity to the different properties mentioned before. 
        \begin{table}
            \caption{Maximum allowed absolute shifts of input properties required for DR3.}
            \label{tab:sensitivity}
            \centering 
                \begin{tabular}{l r r r r r} 
                \hline\hline
                Property & Fid. & Total & Meas. & Detection & Weight \\ 
                & $[10^{-2}]$ & $[10^{-2}]$ & $[10^{-2}]$ & $[10^{-2}]$ & $[10^{-2}]$ \\
                \hline 
                \multicolumn{6}{c}{Ellipticity $\epsilon$} \\
                \hline 
                Mean & 31.56 & 1.07 & 0.83 & 1.89 & 2.76 \\
                Std & 15.84 & 0.78 & 0.72 & 2.26 & 2.44\\
                Kurtosis & $-$62.58 & 9.92 & 30.59 & 18.33 & 121.03\\
                \hline 
                \multicolumn{6}{c}{Half-light radius [pixel]} \\
                \hline 
                Mean & 275.43 & 3.28 & 7.81 & 10.67 & 11.08 \\
                Std & 180.30 & 2.35 & 3.54 & 8.91 & 159.12\\
                \hline 
                \multicolumn{6}{c}{\sersic index} \\
                \hline
                Mean & 142.91 & 1.20 & 0.99 & 6.02 & 38.32 \\
                Std & 117.32 & 2.57 & 1.72 & 6.34 & 38.15 \\
                \hline 
                \end{tabular}
            \tablefoot{This table includes galaxies with measured magnitudes between $20.5$ and $24.5$. The fiducial value of each moment is given as `Fid.', while the allowed shift for the measurement bias alone is denoted as `Meas.'.}
        \end{table}
    
    The results are shown in Table \ref{tab:sensitivity}. Each entry denotes the maximum absolute shift in the respective property that is allowed with the given error budget of the multiplicative bias. 
    We observe that the size changes of objects mostly affect the measurement bias since the relative size to the PSF changes. On the other hand, changes of the ellipticity distribution mostly affect the detection bias since round sources are easier to detect than elongated ones. We can see in Fig.\thinspace\ref{fig:ellipticity_shift} that the detection bias is affected in the opposite direction than the measurement bias and the weighting bias for the shift in absolute ellipticity. That is why the total allowed shift is larger than the shift that would be allowed solely by the measurement bias. For the half-light radius shift, all biases are affected in the same direction, which yields total constraints tighter than the individual ones. In the next section, we compare our results with Table \ref{tab:sensitivity} to evaluate if the approach lined out in this work is sufficient to reach the requirements. 

\section{\label{sc:Results}Results}
    In this section we compare two different observed data sets each with three simulated datasets. To ensure consistency, we define the \Euclid wide-like stack of the COSMOS field as `EWL-COSMOS', and the \Euclid deep-like stack of the COSMOS field as `EDL-COSMOS'. Furthermore, we refer to our fiducial simulation based solely on the Flagship catalogue as `Flagship'. The two simulations based on the augmented morphology are called `Single-\sersic' and `Double-\sersic'. 
    
    To show that the presented augmentation technique is necessary, we also executed simulations based on the morphology that is in the Flagship catalogue. We use the same detection and model fit methods that we also applied to the real data to measure our simulations. Then we compare the core parameters between the measured simulated data and the real data.
    \begin{figure*}
        \centering
        \resizebox{\hsize}{!}{\includegraphics{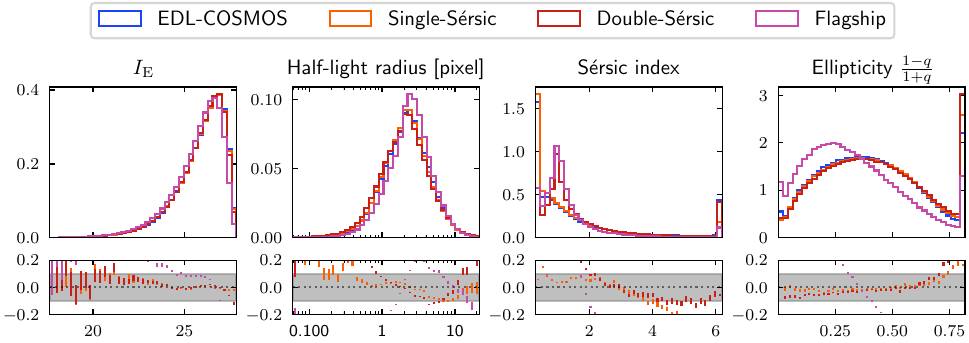}}
        \caption{Comparison of core parameters between real data and image simulations in normalized histograms. The lower panel of each subplot shows the relative difference between image simulations and real data for each of the histogram bins. The image simulations are either based on the copulae or on the Flagship galaxy mock catalogue directly.}
        \label{fig:motivation}
    \end{figure*}
    This can be seen in Fig.\thinspace\ref{fig:motivation}. The disagreement in the ellipticity distribution is the most worrying for the shear calibration. A large part of the difference is driven by faint galaxies below the proposed limit of the \Euclid weak lensing sample of $\SI{24.5}{mag}$. Still these galaxies blend with galaxies from the lensing sample, which requires us to get also their morphology right. Hence, we need to improve the agreement of the calibration simulations with the observational data to accurately calibrate the shear bias. It is already evident in Fig.\thinspace\ref{fig:motivation} that the two trained copula models reproduce the sample properties of morphological parameters better than the simulations based on the Flagship catalogue. The double-\sersic model also shows the peak in the measured \sersic indices at $\nSersic=1$, which is because the \sersic index of the disk is fixed to one as in the Flagship catalogue.
        
    \subsection{\label{sec:selfcal_comparison}EDL-COSMOS sample}
\begin{figure}
            \centering
            \resizebox{\hsize}{!}{\includegraphics{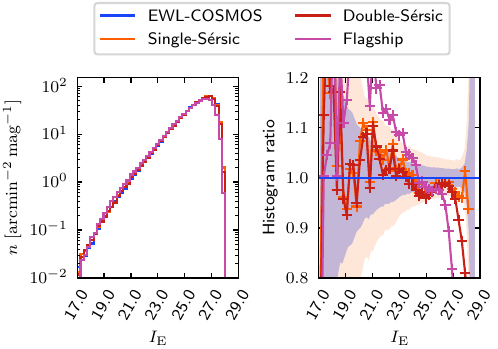}}
            \caption{Number density of galaxies for the EDL-COSMOS field and the simulations. The left panel shows the histogram, while the right panel shows the ratio between the simulations and the reference data in each magnitude bin. The blue and orange shaded regions depict the $1\,\sigma$ and $2\,\sigma$ margins expected from sample variance.}
            \label{fig:number_density}
        \end{figure}

        We want to compare now how well we can reproduce the observed properties of the real data with our simulations. Therefore, we first compared the absolute number density of galaxies in Fig.\thinspace\ref{fig:number_density}. 
        For this figure we also estimated the expected variance in the number density coming from the limited size of the EDL-COSMOS field. To do so we looked at the magnitude histograms of 100 different lines of sight in the Flagship catalogue with an area of $\SI{2}{deg\squared}$. We then applied the completeness determined on our simulations to that magnitude distribution via a random draw. Hence if we only re-detected 50\% of the galaxies in our simulations, we removed galaxies at that magnitude with a 50\% probability from the sample. Then we calculated the mean and the standard deviation of the number counts in each magnitude bin for the 100 lines of sight. In Fig.\thinspace\ref{fig:number_density} we depict the $1\,\sigma$ and $2\,\sigma$ relative deviations from the mean number count. 
        There is a very good agreement until the incompleteness sets in at around 26th magnitude and the difference becomes noisier. Since this paper is aiming at the shear calibration of the EWS, an agreement until this magnitude is already sufficient. 
        
\begin{figure*}
            \centering
            \resizebox{\hsize}{!}{\includegraphics{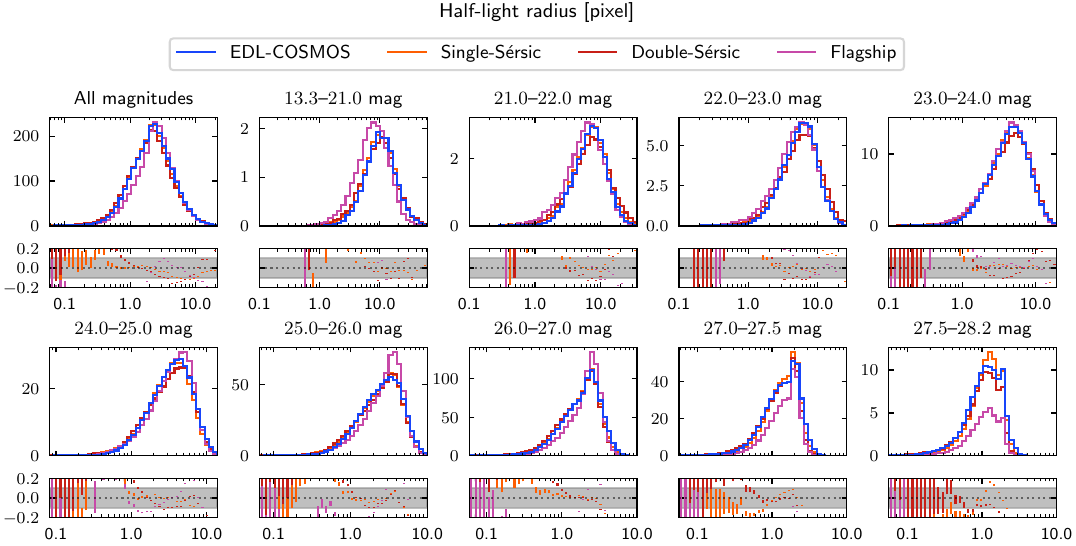}}
            \caption{Half-light radius comparison in arbitrary magnitude bins in comparison to the EDL-COSMOS observations. The upper panels show the augmented and Flagship simulations compared to the EDL-COSMOS data. The lower panels show the relative difference between the histograms and the reference sample. The error bars depict the Poisson uncertainty due to the number of galaxies in each bin. In gray we show the 10\% deviation from the EDL-COSMOS data. The histograms are normalized to the observed area in $\si{arcmin\squared}$ and the bin width.}
            \label{fig:hlr_deep}
        \end{figure*}

\begin{figure*}
            \centering
            \resizebox{\hsize}{!}{\includegraphics{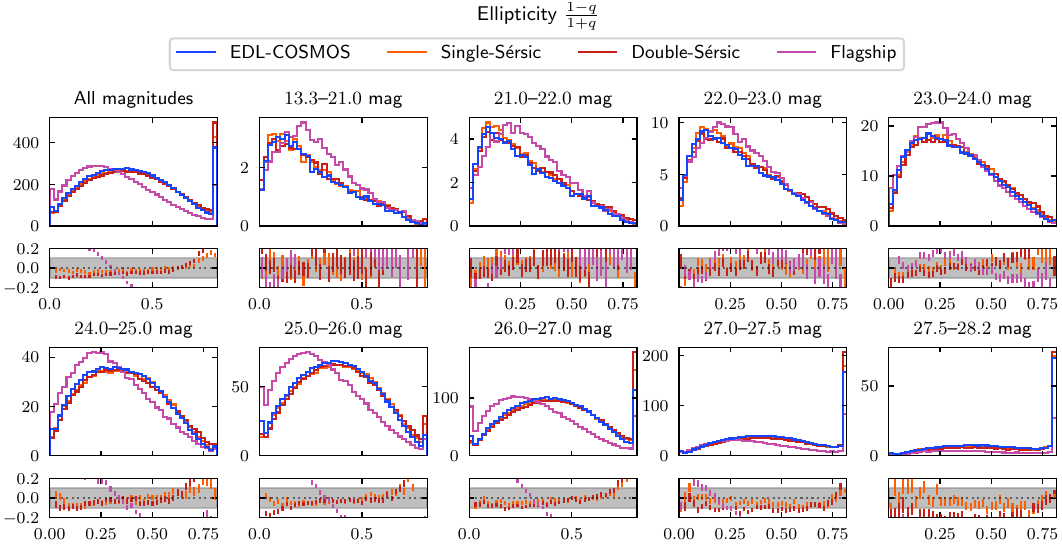}}
            \caption{Ellipticity comparison in arbitrary magnitude bins in comparison to the real data using the EDL-COSMOS observations. The upper panels show the augmented and Flagship simulations compared to the EDL-COSMOS data. The lower panels show the relative difference between the histograms and the reference sample. The error bars depict the Poisson uncertainty due to the number of galaxies in each bin. In gray we show the 10\% deviation from the EDL-COSMOS data. The histograms are normalized to the observed area in $\si{arcmin\squared}$ and the bin width.}
            \label{fig:ellipticity_deep}
        \end{figure*}

        Since we learned the copulae in bins of magnitude, we also want to study the agreement with the real data in the different magnitude bins. We show the histograms for the half-light radii in Fig.\thinspace\ref{fig:hlr_deep} and the histograms for the ellipticities in Fig.\thinspace\ref{fig:ellipticity_deep}. We choose to omit the \sersic index as it is the noisiest quantity and refer to Fig.\thinspace\ref{fig:motivation} for the distribution over the whole sample. In these figures we compare the agreement between single- and double-\sersic models in the simulations with the real data. The lower subplot of each magnitude bin shows the relative deviation of the simulated models to the reference sample, which is the observed data. The gray shaded area marks ten percent of deviation. Each deviation point is depicted as an error bar, which accounts for the Poisson errors of the histogram. 
        It becomes evident that we can reproduce the morphological properties of galaxies in all our chosen magnitude bins very well. 
        
        Overall, we see that we can match the total number density as well as the morphological properties of these galaxies with our approach. We study also the clustering in Appendix \ref{app:clustering} as a consistency check since we assumed it to be correctly modelled in the Flagship simulation. 
        
    \subsection{\label{sec:wide_comparison}EWL-COSMOS sample}
\begin{figure}
            \centering
            \resizebox{\hsize}{!}{\includegraphics{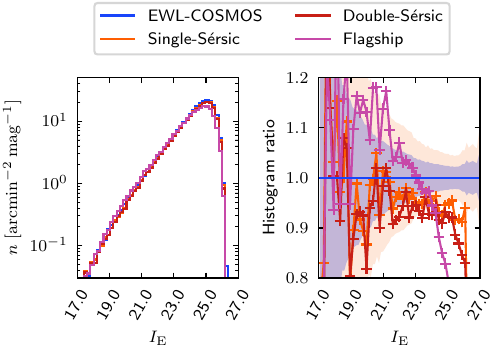}}
            \caption{Same as Fig.~\ref{fig:number_density} but for the EWL-COSMOS data.}
            \label{fig:mag_auto_wide}
        \end{figure}

\begin{figure*}
            \centering
            \resizebox{\hsize}{!}{\includegraphics{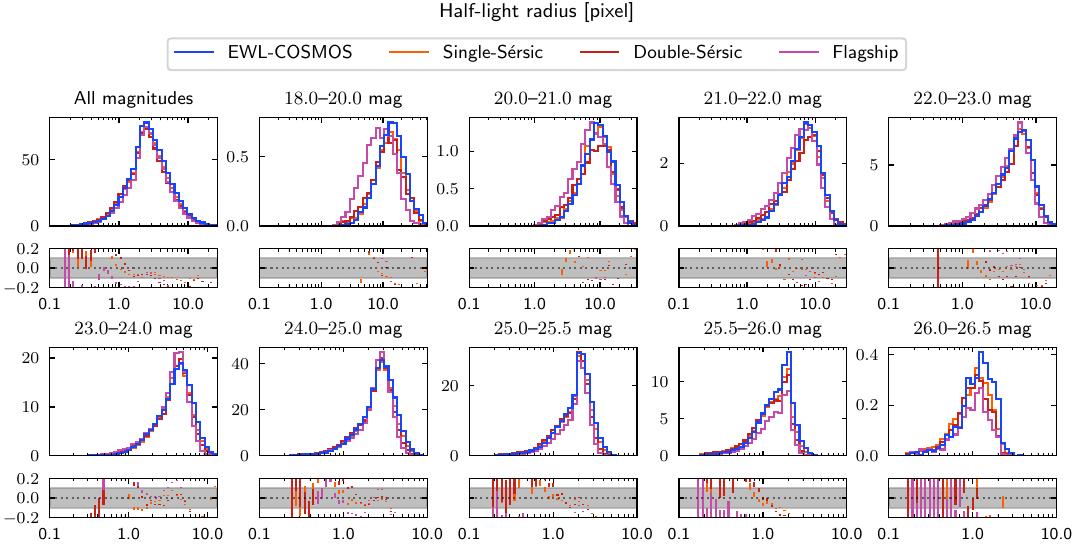}}
            \caption{Same as Fig.~\ref{fig:hlr_deep} but for the EWL-COSMOS data.}
            \label{fig:hlr_wide}
        \end{figure*}

\begin{figure*}
            \centering
            \resizebox{\hsize}{!}{\includegraphics{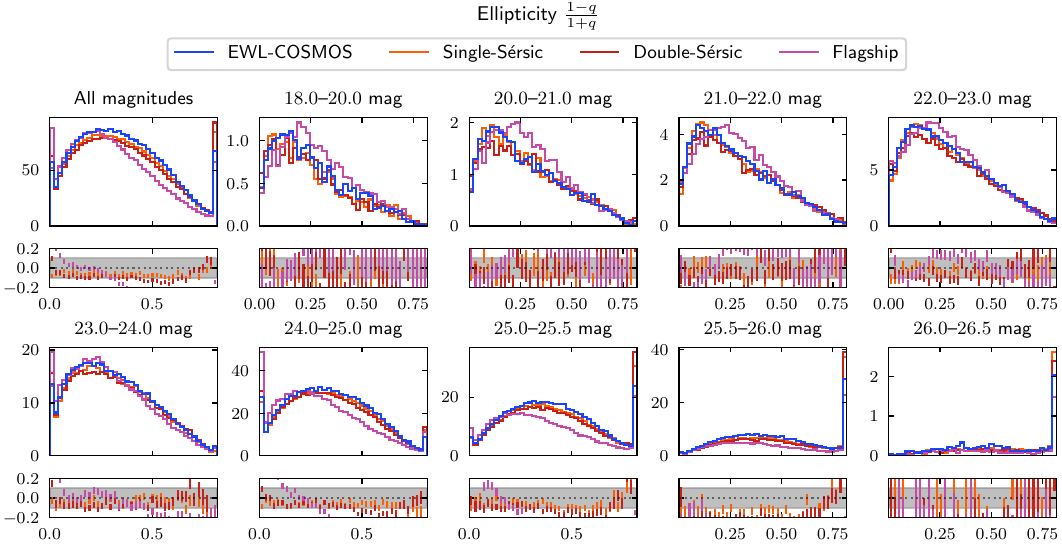}}
            \caption{Same as Fig.~\ref{fig:ellipticity_deep} but for the EWL-COSMOS data.}
            \label{fig:ellipticity_wide}
        \end{figure*}

        For consistency checks we used the input sample that we learned on the deep observations to mimic a normal wide survey observation. As a reference sample we use the wide observation of a region of the COSMOS field that we extracted in Sect.\thinspace\ref{sc:Reference}. Again we show the absolute number density comparison in Fig.\thinspace\ref{fig:mag_auto_wide}, the half-light radius comparison in Fig.\thinspace\ref{fig:hlr_wide}, and the ellipticity comparison in Fig.\thinspace\ref{fig:ellipticity_wide}. Since these observations are shallower, we changed the binning for the comparison. To match the absolute number density, we needed to increase the noise level by 10\% because our simulations do not include depth variations due to masks in individual exposures. The deep observations are less affected by individual masks, but for the wide-like simulations it is a substantial effect. We still observe up to $10\%$ fewer galaxies in the simulations than we find in the data. This can be attributed to substructure getting de-blended in the real data. The simulations include only parametric morphology, which does not lead to fragmentation of bright objects. 

        \begin{table}
            \caption{Summary statistics of morphological distributions for real data and simulations.}
            \label{tab:summary_statistics}
            \centering 
                \begin{tabular}{l r r r r} 
                \hline\hline
                Statistic & COSMOS & $\Delta$Single & $\Delta$Double & $\Delta$Flagship\\
                & (EWL) & $[10^{-2}]$ & $[10^{-2}]$ & $[10^{-2}]$ \\
                \hline 
                \multicolumn{5}{c}{Ellipticity $\epsilon$} \\
                \hline 
                Mean & 0.314 & $-$0.08 & 0.02 & $-$2.27\\
                Std & 0.190 & $-$0.19 & 0.27 & $-$0.53 \\
                Kurtosis & $-$0.640 & $-$3.46 & $-$2.09 & 21.67 \\
                \hline 
                \multicolumn{5}{c}{Half-light radius [pixel]} \\
                \hline 
                Mean & 4.331 & $-$14.29 & $-$26.79 & $-$44.94\\
                Std & 2.210 & $-$8.71 & $-$11.93 & $-$31.91\\
                \hline 
                \multicolumn{5}{c}{\sersic index} \\
                \hline 
                Mean & 1.142 & $-$8.29 & 5.39 & $-$4.89 \\
                Std & 0.781 & $-$7.84 & $-$23.97 & $-$37.83 \\
                \hline
                \end{tabular}
            \tablefoot{We list here the reference values measured on the real EWL-COSMOS observations together with the relative difference between each simulation and the reference. For the statistics we limited the sources again following $20.5<\IE<24.5$.}
        \end{table}
        In Table \ref{tab:summary_statistics} we list the summary statistics for which we determined the requirements in Sect.\thinspace\ref{sc:sensitivity}. Here we limited the sample to galaxies between $20.5$ and $24.5$ magnitude since those sources are most relevant for the weak lensing analysis. Comparing with Table \ref{tab:sensitivity}, we can see that the requirements for the ellipticity statistics are already all fulfilled. The mean of the half-light radius deviates still more than the requirement would allow and also the mean of the \sersic index deviates from the requirement. These two morphological parameters are very correlated with each other and also with the PSF model. Therefore, it is likely that the more accurate PSF modelling of the actual ground segment simulations yields a better agreement than the \texttt{PSFex} model that was used in this work. 
        
\begin{figure*}
            \centering
            \resizebox{\hsize}{!}{\includegraphics{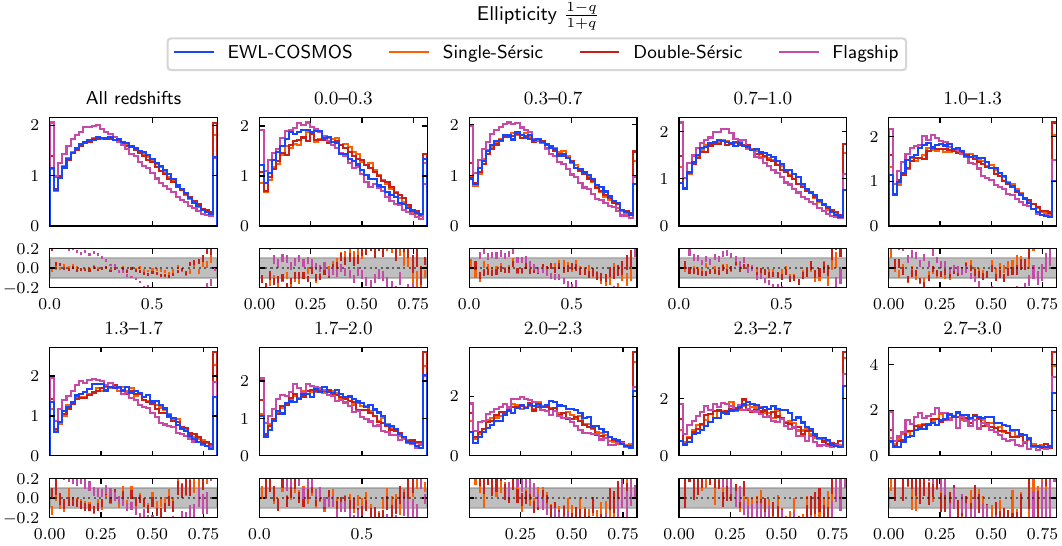}}
            \caption{Ellipticity comparison in arbitrary redshift bins in comparison to the real data using the EWL-COSMOS observations. The upper panels show the augmented and Flagship simulations compared to the EWL-COSMOS data. The lower panels show the relative difference between the histograms and the reference sample. The error bars depict the Poisson uncertainty due to the number of galaxies in each bin. In gray we show the 10\% deviation from the EWL-COSMOS data. The histograms are normalized as a probability density.}
            \label{fig:ellipticity_wide_redshift}
        \end{figure*}

        We find that our setup indeed seems robust and can be used to simulate a wide-survey-like mosaic with the learned input from deeper observations. In particular we manage to capture the ellipticity distribution at the faint end very well, which is most important for the shear bias calibration together with the sizes of objects. We can not test the agreement within tomographic bins, since we only simulated the VIS channel. In Fig.\thinspace\ref{fig:ellipticity_wide_redshift} we show the ellipticity distributions binned in redshift, where we used the high-quality photometric redshifts for the EWL-COSMOS and the input redshifts from our image simulations for the binning. 
        We see a good agreement between data and simulations also here, although there seems to be no major trend with redshift. Hence, we conclude that the copulae capture also most of the redshift dependence, which enables a tomographic calibration in the future.

    \subsection{\label{sec:ShearBias}Shear bias}
        To quantify the impact that these adapted morphologies have in comparison to using a simulation based on the Flagship galaxy catalogue, we generated image simulations of both and measure the shear bias of each in the following. We make use of both shape- and pixel noise cancellation and use large input shears up to $0.2$ for a faster estimate. Since we are only interested in the residual shear bias, we do not need to care about eventual non-linearities of the shear measurement method \citep[see, e.g.,~Appendix A of][]{Schrabback25}. The resulting shear biases can be seen in Table \ref{tab:shear_biases}.
            \begin{table}
                \caption{Relative shear biases for the different image simulations.}
                \label{tab:shear_biases}
                \centering 
                \begin{tabular}{l r r r r} 
                \hline\hline
                Method & $\Delta\mbias{\mathrm{total}}$ & $\Delta\mbias{\mathrm{meas}}$ & $\Delta\mbias{\mathrm{det}}$ & $\Delta\mbias{\mathrm{weight}}$ \\ 
                & $[10^{-2}]$ & $[10^{-2}]$ & $[10^{-2}]$ & $[10^{-2}]$ \\
                \hline
                Single-\sersic & 1.14 & 1.23 & $-$0.20 & 0.03 \\
                Double-\sersic & $-$0.74 & $-$0.89 & $-$0.06 & 0.15 \\
                \hline
                \end{tabular}
                \tablefoot{This table includes galaxies with measured magnitudes between $20.5$ and $24.5$. We quote here the relative bias with respect to a simulation based on the Flagship catalogue.}
            \end{table}
        We list them as a relative shift with respect to the Flagship-based image simulations. Additionally, we split the bias components up into contributions caused by the measurement algorithm itself, the pure detection bias, and the bias that is introduced by the shear weights. It is evident that we get percent-level differences in the multiplicative shear bias depending on the morphology used. The statistical uncertainty on individual bias estimates is $2\times10^{-3}$. Hence, the uncertainty on the difference between two estimates neglecting the correlation between the two is roughly $3\times10^{-3}$. Therefore, the relative bias differences compared to the Flagship simulation are at least $2\,\sigma$ significant. Even the choice of the parametric model for the copulae has a substantial impact on the biases. In particular the measurement bias seems to be sensitive to the chosen model, although both of them agree in their measured morphology with the observed data. A sensitivity study of \textsc{Metacalibration} to the chosen model is an interesting endeavour for future work. Since no model is implicitly assumed by \textsc{Metacalibration}, it is potentially less sensitive to the model chosen for the image simulations.

\section{\label{sc:Conclusion}Conclusions}
    In this work we developed the framework to implement observed galaxy morphology distributions into the image simulations required for the shear calibration. Vine copula models that allow for conditional sampling can be used to generate realistic galaxy properties as a function of magnitude and redshift. Assuming that these properties are correctly modelled in the Flagship galaxy mock catalogue, we were able to generate a new simulation input sample. The outputs of image simulations based on this new catalogue resemble the real data better than the simulations based on the original Flagship catalogue, and therefore validate our approach. 

    One crucial part of this work was to take into account observational features like the incompleteness. We treated this with a combination of a magnitude weighting and a binary classifier, which learns the detection probability of a galaxy. This treatment had to be applied considering also the exposure time variations that come with the observed data. Further we showed that the excellent match we produced for the deep reference sample can be used to generate wide-survey-like image simulations, which also agree with the observed data. 
    
    With dedicated simulations, we propagated the required accuracy of the multiplicative shear bias into requirements on key morphological parameter distributions. Comparing our results with these requirements, we find that we can already reproduce the ellipticity distribution of the observed data within final mission requirements. Only the requirements for the mean of the half-light radius and the \sersic index cannot yet be met, but might be solved with a more accurate PSF model. However, it is worth mentioning that these requirements were derived for the full survey area, which will only be available in the final data release. Considering the scaling between the first and the final data release, we expect about a factor three lower requirement on the multiplicative bias for the first data release. With the relaxed requirements, we are close to meeting also the required accuracy of the half-light radius. In any case, the approach outlined in this work produces more realistic image simulations than using the Flagship catalogue directly. 

    When analysing the shear biases that arise from the different morphology models, we find percent-level shifts in the multiplicative bias, which are not acceptable for the cosmological analysis of \Euclid data. Therefore, this work is crucial to mitigate a systematic bias of the multiplicative shear bias due to wrong galaxy morphology distributions in the image simulations. For example, the sensitivity to the morphological parameters of the bulge, found in \citet{EP-Congedo}, can be addressed with this work by matching the simulation inputs to the observed distribution of bulge sizes and \sersic indices. The decision between a single-\sersic and a double-\sersic model is not obvious, since observationally there is no reason to believe that one of them is more likely than the other. Only the better agreement of the measured \sersic index would suggest using a single-\sersic model. Since the observed difference is mostly caused by a measurement bias, a future measurement using \textsc{Metacalibration} will show if a model bias of \textsc{LensMC} could be causing the difference. For the final calibration of the first data release of \Euclid we aim to study the sensitivity to the chosen model of both shape measurement algorithms and derive a final decision from that. 

    Finally, this framework will be used to update the input catalogues for the large-scale image simulations that will be used to deliver the shear calibration for the first data release of \Euclid. For the final data release, the reference area shall be expanded to at least $\SI{5}{\deg\squared}$. Therefore, the other auxiliary fields of \Euclid described in \cite{Scaramella-EP1} shall be included in the analysis. For the first data release, we are aiming to use only the COSMOS field due to its high-quality photometric redshifts. The increase in area is mainly needed for a more accurate calibration of the number counts of galaxies, while key morphological parameters like the ellipticity can already be constrained with smaller fields like COSMOS used in this analysis. Furthermore, the agreement has to be checked also in the tomographic redshift bins that enter the cosmological analysis. Since we did image simulations only of the VIS channel, we could not measure photometric redshifts ourselves. We showed that a binning in input redshifts shows good agreement between data and simulations and follow that a calibration in tomographic redshift bins will be possible in the future. Hence, this work outlines the technical details of one of the major steps in the shear calibration plan, which is one of the keys to unlock the huge scientific potential of \Euclid.
%
%

\begin{acknowledgements}
The Innsbruck authors acknowledge support provided by
the Austrian Research Promotion Agency (FFG) and the Federal Ministry of
the Republic of Austria for Innovation,  Mobility, and Infrastructure (BMIMI) via the Austrian Space Applications Programme with grant numbers 899537, 900565, 911971, and 928759. This research was funded in whole or in part by the Austrian Science Fund (FWF) 10.55776/F101300. NM acknowledges the funding of the French Agence Nationale de la Recherche for the PISCO project (grant ANR-22-CE31-0004). HH acknowledges support from the European Research Council (ERC) under the European Union’s Horizon 2020 research and innovation programme with Grant agreement No. 101053992. GC thanks the United Kingdom Space Agency for support.
\AckEC
\AckCosmoHub

\end{acknowledgements}

%
%

\bibliography{my, Euclid}

%

\begin{appendix}
  \onecolumn 

\section{Star-galaxy separation}\label{app:s_g}
\begin{figure}
        \centering
        \resizebox{\hsize}{!}{\includegraphics{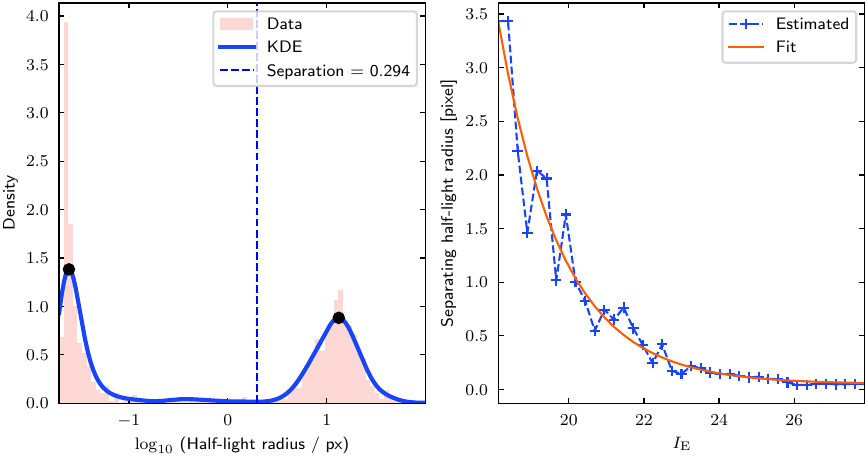}}
        \caption{Star-galaxy separation based on the fitted (PSF-corrected) half-light radius. The left panel shows the bimodal distribution in the magnitude bin around magnitude 19. The right panel shows the fitted behaviour to the separation cut as a function of magnitude.}
        \label{fig:s_g_sep}
    \end{figure}

    To separate galaxies from stars, we adopted a cut in the fitted (PSF-corrected) half-light radius of sources. This half-light radius is therefore deconvolved with the PSF. In the left panel of Fig\thinspace\ref{fig:s_g_sep} we show the procedure exemplary for bright sources around magnitude 19. 
    We fit a KDE to the distribution of measured half-light radii in logarithmic space. This distribution is clearly bimodal, where the population of small half-light radii is caused by stars. We determined the two dominant peaks of the KDE and put a separation cut at the lowest density between those two. This separation cut depends on the magnitude of the sources. We modeled this dependency by determining the separation cut in 40 evenly spaced magnitude bins between magnitude 18 and 28. In the right panel of Fig.\thinspace\ref{fig:s_g_sep}, we show this fitted magnitude dependency. We validated this star-galaxy separation by comparing the number counts of sources classified as stars with the stellar population model for the COSMOS field we used in Sect.\thinspace\ref{sc:Simulations}. We found an excellent agreement with the model until 24th magnitude.

\section{Ellipticity pre-processing}{\label{app:ellipticity}}
    As it can be seen in Fig.\thinspace\ref{fig:ellipticity_deep}, fitting without priors produces ellipticity spikes at the extrema for galaxies fainter than $\IE>25$. For these faint galaxies, it is not required to perfectly re-produce the $p(\epsilon)$, but we need to remove the spike for practical reasons in the copula training. To ease the training of the copulae, we re-distribute fitted ellipticities that lie very close to the fit boundaries. We make use of a KDE with variable bandwidth, where for each point the bandwidth is given as the measurement uncertainty. Since we would find probability mass outside of the boundaries of the axis ratio, we used truncated Gaussian kernels to constrain the KDE to the physically meaningful values. 
    \begin{figure}
        \centering
        \resizebox{0.5\hsize}{!}{\includegraphics{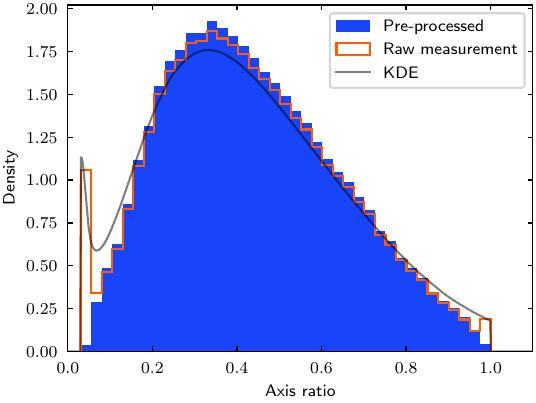}}
        \caption{KDE based pre-processing of ellipticity distributions for $26.5<\IE<26.75$.}
        \label{fig:ellip_preproc}
    \end{figure}
    We visualize the procedure in Fig.\thinspace\ref{fig:ellip_preproc}. The orange histogram shows the spikes at the extreme axis ratios for galaxies with magnitudes between 26.5 and 26.75. After re-distribution the axis ratios follow the blue distribution, which is spike-free. 
    
\section{Clustering}\label{app:clustering}
    Another important statistic that the simulations need to match for an accurate bias calibration is the clustering of galaxies \citep{Martinet-EP4}. We defined two first-order clustering statistics, which give an idea of the degree of agreement: the distance to the nearest neighbour and the number of neighbours in a circle around the object. We did this comparison in seven arbitrarily chosen magnitude bins and for the full sample. The detection in both cases was run with exactly the same settings. Only galaxies in the same magnitude bin are accounted for as possible neighbours. 
    \begin{figure*}
        \centering
        \resizebox{\hsize}{!}{\includegraphics{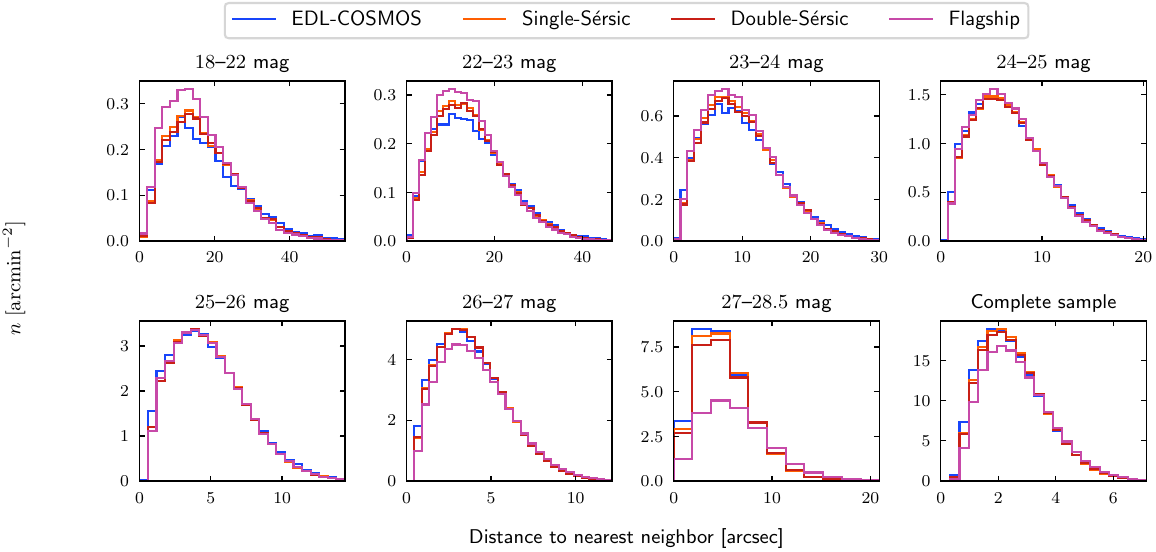}}
        \caption{Distance to the nearest neighbour for the simulations and the real data. Each panel illustrates one arbitrarily chosen magnitude bin.}
        \label{fig:distance_nn}
    \end{figure*}
    \begin{figure*}
        \centering
        \resizebox{\hsize}{!}{\includegraphics{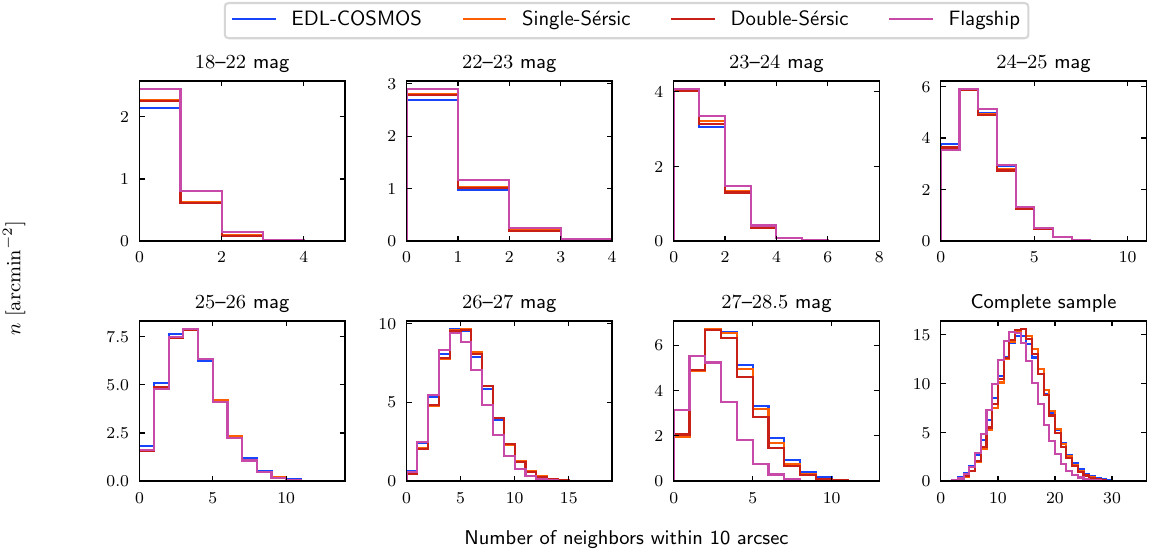}}
        \caption{Number of neighbours within a circle of \ang{;;10}. Each panel illustrates one arbitrarily chosen magnitude bin.}
        \label{fig:n_neighbors}
    \end{figure*}
    In Fig.\thinspace\ref{fig:distance_nn} we show the distance to the nearest neighbour of the same magnitude bin and in Fig.\thinspace\ref{fig:n_neighbors} the number of neighbours of the same magnitude bin within \ang{;;10}.
    In general we see a good agreement in the individual magnitude bins and for the complete sample. Hence, we conclude that our image simulations reproduce a realistic clustering. 

\end{appendix}

\end{document}